\documentclass[]{jfm}
\usepackage{graphicx}
\usepackage{epstopdf,epsfig}
\AppendGraphicsExtensions{.tiff}
\usepackage{float}
\usepackage{newtxtext}
\usepackage{newtxmath}
\usepackage{natbib}
\usepackage[colorlinks]{hyperref}
\usepackage{etoolbox}
\usepackage[noabbrev]{cleveref}
\usepackage{bbm}
\hypersetup{
	colorlinks = true,
	linkcolor=blue,
	urlcolor = blue,
	citecolor = blue,
}
\usepackage{empheq}

\makeatletter

\usepackage[dvipsnames]{xcolor}

\usepackage{fancyhdr}
\usepackage{tkz-euclide}
\usetikzlibrary{shapes.callouts,arrows.meta,calc,shadings}
\usetikzlibrary{math}

\usetikzlibrary{external}
\tikzset{external/system call={pdflatex \tikzexternalcheckshellescape 
		-halt-on-error
		-interaction=batchmode 
		-jobname "\image" "\texsource"
		&& pdftops -eps "\image.pdf"}}
\patchcmd{\NAT@citex}
{\@citea\NAT@hyper@{%
		\NAT@nmfmt{\NAT@nm}%
		\hyper@natlinkbreak{\NAT@aysep\NAT@spacechar}{\@citeb\@extra@b@citeb}%
		\NAT@date}}
{\@citea\NAT@nmfmt{\NAT@nm}%
	\NAT@aysep\NAT@spacechar\NAT@hyper@{\NAT@date}}{}{}

\patchcmd{\NAT@citex}
{\@citea\NAT@hyper@{%
		\NAT@nmfmt{\NAT@nm}%
		\hyper@natlinkbreak{\NAT@spacechar\NAT@@open\if*#1*\else#1\NAT@spacechar\fi}%
		{\@citeb\@extra@b@citeb}%
		\NAT@date}}
{\@citea\NAT@nmfmt{\NAT@nm}%
	\NAT@spacechar\NAT@@open\if*#1*\else#1\NAT@spacechar\fi\NAT@hyper@{\NAT@date}}
{}{}

\makeatother

\defcitealias{yang2014effect}{2014\textit{a}}
\defcitealias{yang2014large}{\textit{b}}

\usepackage{layouts}
\usepackage{scalerel,stackengine}
\stackMath
\newcommand\reallywidehat[1]{%
	\savestack{\tmpbox}{\stretchto{%
			\scaleto{%
				\scalerel*[\widthof{\ensuremath{#1}}]{\kern-.6pt\bigwedge\kern-.6pt}%
				{\rule[-\textheight/2]{1ex}{\textheight}}
			}{\textheight}%
		}{0.5ex}}%
	\stackon[1pt]{#1}{\tmpbox}%
}
\usepackage{quoting,xparse}
\usepackage{bm}
\usepackage{setspace}

\newcommand{\RomanNumeralCaps}[1]
\linenumbers

\usepackage{epstopdf}
\AppendGraphicsExtensions{.tiff}

\shorttitle{Initial stage of wave generation}
\shortauthor{T. Li and L. Shen}

\title{A theoretical study of the upper bound of surface elevation variance in the Phillips initial stage during wind-wave generation}

\author{Tianyi Li\aff{1,}\aff{2}
	\and Lian Shen\aff{1}\corresp{\email{shen@umn.edu}}}

\affiliation{\aff{1}{Department of Mechanical Engineering and St. Anthony Falls Laboratory, University of Minnesota,
	Minneapolis, MN 55455, USA}
 \aff{2}{Atmosphere, Earth and Energy Division, Lawrence Livermore National Laboratory, Livermore, CA 94550, USA}}

\begin{document}
	\maketitle
	
\begin{abstract}
The resonance mechanism in the initial stage of wind-wave generation proposed by Phillips (\textit{J. Fluid Mech.}, vol. 2, 1957, 417--445) is a foundation of wind-wave generation theory, but a precise theoretical quantification of wave energy growth in this initial stage has not been obtained yet after more than six decades of research. In this study, we aim to address this knowledge gap by developing an analytical approach based on a novel complex analysis method to theoretically investigate the temporal evolution of the wave energy in the Phillips initial stage. We quantitatively derive and analyse the growth behaviour of the surface wave energy and obtain an analytical solution for its upper bound. Our result highlights the crucial effects of surface tension. Because the phase velocity of gravity--capillary waves has a minimal value at a critical wavenumber, gravity--capillary waves and gravity waves (which neglect surface tension) exhibit distinct resonance curve properties and wave energy growth behaviours. For gravity waves, the resonance curve extends indefinitely; for gravity--capillary waves, it either forms a finite-length curve or does not exist, depending on the wind speed. The leading-order term of the upper-bound solution of the energy of gravity waves increases linearly over time, while for gravity--capillary waves, the term increases linearly over time under strong wind conditions but remains finite under weak wind conditions. This theoretical study provides an analytical framework for the generation of wind-waves in the Phillips initial stage, which may inspire further theoretical, numerical, and experimental research.	
\end{abstract}
	
	\begin{keywords}
		wind--wave interactions, wave--turbulence interactions, gas/liquid flow
	\end{keywords}

\section{Introduction}
The mechanism by which turbulent airflow generates ocean surface waves has been actively researched in recent decades.
Understanding the complex physical processes underlying wind-wave generation and evolution is crucial for ocean modelling, weather forecasting, and naval engineering. 
Many theoretical studies on the wind-wave generation mechanisms have been conducted in the past several decades, including the seminal works by \citet{phillips1957generation} and \citet{miles1957generation}. 

Phillips (\citeyear{phillips1957generation}) proposed a resonance mechanism between surface waves and air turbulence pressure fluctuations that governs the early growth stages of waves. This theory includes two stages in wave development: the initial stage and the principal stage.
The initial stage in Phillips theory accounts for the initial excitation of surface deformation in response to turbulent airflow. In the initial stage, resonance occurs when the wave phase velocity equals the convection velocity of turbulent pressure fluctuations at the water surface. Phillips employed Taylor's frozen hypothesis~\citep{taylor1938spectrum} to model the temporal--spatial evolution of air pressure fluctuations. According to this hypothesis, for air turbulence pressure fluctuations propagating in the same direction as the wind, their spatial--temporal evolution can be modelled as convection motions. When surface waves with specific wavenumbers satisfy the resonance condition, the resonance mechanism leads to quadratic temporal growth in the wave energy at these wavenumbers.
Recently, numerical evidence of the Phillips resonance mechanism in the initial stage was reported by~\citet{li2023direct}.
After the initial stage of wave development, the principal stage occurs, during which the decorrelation effects of turbulent air pressure fluctuations are taken into account; thus, Taylor's frozen hypothesis is no longer applicable in this stage. In the principal stage, the wave energy components increase linearly over time, which has been validated through numerical simulations~\citep{lin2008direct,li2022evolution,li2022principal} and laboratory experiments~\citep{zavadsky2017water,geva2022excitation}. 

The \citet{phillips1957generation} theory is applicable to the early stage of wind-wave generation when the surface deformation is not large enough to alter airflow turbulence.  Later, when the wave evolution and air turbulence are fully coupled, the critical-layer instability mechanism proposed by \citet{miles1957generation} leads to an exponential increase in the surface wave energy. This critical-layer instability occurs when the mean velocity of the airflow at a certain height equals the wave phase velocity, which has been observed in field measurements~(e.g., Hristov, Miller \& Friehe~\citeyear{hristov2003dynamical}). Notably, the non-separation sheltering mechanism proposed by~\citet{belcher1993turbulent} elucidates the complex momentum exchanges between airflow and surface waves when surface waves of finite amplitude are present.   {Recent studies~\citep{zavadsky2017water,geva2022excitation} also discovered linear growth of wave energy at the later nonlinear stage of wind--wave interactions, which can be supported by the nonlinear Phillips' mechanism.}

The above theoretical works established the foundation for the study of wind-wave generation. Benefiting from recent developments in computational capabilities and experimental techniques, wind--wave interactions have been studied through numerical simulations~(e.g., Druzhinin,  Troitskaya \& Zilitinkevich~\citeyear{druzhinin2017study}; Hao~\textit{et al.}~\citeyear{hao2018simulation}; Li \& Shen~\citeyear{li2022principal}, \citeyear{li2023direct}; Cimarelli, Romoli \& Stalio~\citeyear{cimarelli2023wind}; Matsuda~\textit{et al.}~\citeyear{matsuda2023effects}; Zhu~\textit{et al.}~\citeyear{zhu2023moving}) and laboratory and field experiments~\citep[e.g.,][]{liberzon2011experimental,grare2013growth,geva2022excitation,kumar2024spatial}.
However, a complete understanding of the wind-wave generation mechanism has not yet been achieved yet; hence, there is still a critical need to examine previous theoretical studies to improve our understanding.

In the present study, we revisit the classic work of \citet{phillips1957generation}. 
We focus on the Phillips initial stage of wind-wave generation and the associated wave energy growth behaviour. 
The purpose of this study is to address an important knowledge gap regarding the temporal evolution characteristics during the initial stage, which are not fully understood. 
Phillips~(\citeyear{phillips1957generation}) highlighted the significance of the resonance curve in the initial excitation of surface waves. 
The resonance curve represents the wavenumbers at which the wave phase velocity matches the convection velocity of the air pressure fluctuations at the water surface.
The wave energy associated with the wavenumbers on the resonance curve increases quadratically over time, while the wave energy associated with the wavenumbers far from the resonance curve does not increase but instead oscillates over time. 
The temporal increase in the total wave energy, which is represented by the integration of the wave components over all wavenumbers, exhibits more complex behaviours. 
\citet{phillips1957generation} argued that the total wave energy would increase linearly over time in the initial stage of wind-wave generation; however, he did not obtain a quantitative expression for the wave energy, and the corresponding analysis was not mathematically rigorous. 
These important issues have been pointed out in recent papers~\citep{zavadsky2017water,shemer2019evolution,li2023direct}. Another important problem that needs to be addressed is the minimum wind speed required to generate surface waves. 
Phillips~(\citeyear{phillips1957generation}) argued that wind can generate surface waves only when the convection velocity of pressure fluctuations at the water surface exceeds the minimum phase velocity of gravity--capillary waves.  
However, the validation of this minimum wind speed has yet to be established theoretically. 
Moreover, how surface waves respond to airflow when the convection velocity of pressure fluctuations at the water surface is less than the minimum velocity of gravity--capillary waves but finite remains unknown, for which \citet{phillips1957generation} only considered the zero wind convection velocity scenario.

Our study aims to theoretically understand the temporal wave evolution in the initial stage. 
By developing a novel approach based on a complex analysis method, we obtain quantitative analytical results of the upper-bound solution for the temporal growth behaviours of surface wave energy for the first time. 
We emphasize that this work has not yet analytically solved the problem completely. 
But the theoretical solution on the upper bound of wind-wave growth may provide important insights into the physical problem and lead to a rigorous mathematical framework for follow-up analytical and numerical studies and analyses of experimental data.

Our results highlight the significant effects of surface tension on the resonance curve and wave energy growth in the initial stage of wind-wave generation. 
It is well known that the phase velocity of gravity waves (i.e. when the surface tension effect is ignored) increases with increasing wavenumber; however, for gravity--capillary waves (i.e. with the surface tension effect considered), the phase velocity exhibits a minimal value at a critical wavenumber.  We find that this difference results in distinctly different resonance curve properties and wave growth behaviours.
The resonance curve of gravity waves begins on the $k_x$-axis and continues indefinitely in the first quadrant of the wavenumber space $(k_x,k_y)$.
Here, $k_x$ is the wavenumber on the $x$-axis, corresponding to the mean wind direction, and $k_y$ is the wavenumber in the $y$-axis direction.
Conversely, the resonance curve of gravity--capillary waves exhibits distinct characteristics.
When the wind speed is sufficiently high, the convection velocity of pressure fluctuations at the water surface exceeds the minimum phase velocity of the gravity--capillary waves, and the resonance curve displays a finite length.
In the first quadrant of the wavenumber space, the resonance curve begins at a point on the $k_x$-axis and terminates at another point on the same axis.
However, under relatively weak wind conditions, the convection velocity of the pressure fluctuations at the water surface is lower than the minimum phase velocity of the surface gravity--capillary waves, and no resonance curve can be found in the wavenumber space.
Consequently, we demonstrate analytically that for gravity waves, the leading-order term of the upper-bound solution of the wave energy increases linearly over time for all wind speeds.
In contrast, for gravity--capillary waves, the leading-order term of the upper-bound solution of the wave energy increases linearly over time under strong wind conditions, while it remains finite under weak wind conditions.
We present mathematical derivations for both gravity waves and gravity--capillary waves in the subsequent sections.

The remainder of this paper is organised as follows. In \S~\ref{sec2}, we describe the problem and summarise the main results. In \S~\ref{sec3}, we present the temporal evolution analysis of gravity waves when ignoring surface tension effects. 
In \S~\ref{sec4}, we discuss the temporal evolution of gravity--capillary waves considering surface tension effects. The conclusions and discussions are presented in \S~\ref{sec5}.

\section{Problem definition and main results}\label{sec2}
In this section, we first define the problem of wind-wave generation in the initial stage of Phillips theory in \S~\ref{sec2.1}. In \S~\ref{sec2.2}, we review the original approach proposed by Phillips and discuss its limitations. Our new approach is presented in \S~\ref{sec2.3}.

\subsection{Problem definition}\label{sec2.1}
In the early stage of the wind-wave generation process, the surface elevation is not substantially high, such that nonlinear wave--wave interactions can be ignored~\citep{phillips1957generation}.  Therefore, the linearised water wave equation is adequate to model the evolution of surface elevation.
The linearised governing equation for the surface wave elevation $\eta$ can be expressed in wavenumber space as follows:
	\begin{align}
		\frac{\partial^2\hat\eta}{\partial  t^2}+\Lambda^2(\bm k)\hat\eta=-\frac{k}{\rho_w}\hat{p}(\bm k,t),\label{eq:2.1}
	\end{align}
where $(\hat\cdot)$ represents the Fourier transform.  $\hat\eta(\bm k,t)$ and $\hat p(\bm k,t)$ denote the complex-valued wave amplitude and pressure, respectively, at the water surface for wavenumber $\bm k$ and time $t$.  The angular frequency $\Lambda(\bm k)$ indicates the dispersion of the surface waves and is $\Lambda(\bm k)=\sqrt{g|\bm k|}$ for gravity waves and $\Lambda(\bm k)=\sqrt{g|\bm k|+(\sigma/\rho_w)|\bm k|^3}$ for gravity--capillary waves. Here, $g$ denotes gravitational acceleration, $\sigma$ denotes the surface tension of the air--water interface, and $\rho_w$ denotes water density.

 {In the present study, the wind-induced water shear current is not considered. The influence of water shear flow on wave growth has been investigated in several studies. \cite{nove2020effect} demonstrated that the presence of longitudinal water currents generates fine-scale surface wrinkles. \cite{geva2022excitation} derived coupled Orr--Sommerfeld equations to include the effects of water shear flow. In the present study, we assume the water is initially calm and the air--water interface is flat, consistent with the framework established by \citet{phillips1957generation}.  While wind shear can induce drift velocities in the water as waves develop, previous numerical studies have shown that, during the principal stage of wind-wave generation starting from calm water, the amplitude of the drift current velocity is small compared with the wave phase speed and can be neglected~\citep{li2022principal}.  Our present study focuses on the initial stage of wave generation, which precedes the principal stage in Phillips theory. At this stage, the assumption of zero surface drift velocity is reasonable.  The assumption of negligible surface drift velocity has also been adopted in other literatures with a focus on wave growth when wind-induced water currents are not strong (e.g., Perrard \textit{et al.} \citeyear{perrard2019turbulent}; Zhang, Hector \& Moisy \citeyear{zhang2023wind}; Maleewong \& Grimshaw \citeyear{maleewong2022evolution}).
The scenario of appreciable shear currents studied by \citet{nove2020effect} and \citet{geva2022excitation} is also an important problem, but is beyond the scope of the present work and will be a subject of our future research.}
 {In the present study, water dissipation is not considered, which is a mild assumption and is consistent with Phillips’ theory.  During the initial stage of wind-wave generation, when waves are excited from a calm water surface, the wind input dominates and overcomes water dissipation. The effect of water viscosity can be incorporated into the governing equation~\eqref{eq:2.1} by introducing an additional term $4\nu k^2 \partial_t \eta$~\citep{perrard2019turbulent}, where $\nu$ is water kinematic viscosity.
Our future research will address wind-wave generation considering the effects of water viscosity, which is particularly relevant in the presence of sheared water currents.}

In the theory presented in \citet{phillips1957generation}, the temporal evolution of air pressure fluctuations is described according to Taylor's frozen hypothesis~\citep{taylor1938spectrum}, which can be formulated as 
	\begin{align}
		\hat{p}(\bm k,t)=\hat{p}_0(\bm k)\exp(- \mathrm{i}\bm k\cdot \bm U t),
	\end{align}
where $\hat{p}_0(\bm k)$ denotes the air pressure fluctuations at time $t=0$, and $\bm U$ denotes the convection velocity of the air pressure fluctuations.  
In the present study, we assume that $\bm U$ is a constant vector pointing in the $x$-direction with a magnitude of $U$.
The convection velocity $U$ in our study represents the convection of pressure fluctuations at the air--water interface. As demonstrated by Li \& Shen (\citeyear{li2022principal}, figure 8), the convection velocity of these air pressure fluctuations at the water surface exhibits only slight variations with respect to the wavenumbers over the range of scales relevant to wind-wave generation. Therefore, assuming a constant convection velocity of air pressure fluctuations at the air--water interface across different wavenumbers is a reasonable simplification for this analysis. This assumption aligns with previous theoretical studies of wind--wave interactions~\cite[e.g.,][]{perrard2019turbulent}.
Under this assumption, the governing equation for $\hat\eta$ becomes
	\begin{align}
		\frac{\partial^2\hat\eta}{\partial  t^2}+\Lambda^2(\bm k)\hat\eta=-k\hat{p}_0(\bm k)\exp(- \mathrm{i}\bm k\cdot \bm U t).
	\end{align}
 {The assumption that turbulent pressure fluctuations retain coherence over short durations is based on Taylor’s frozen hypothesis, which is applicable in the initial stage of wind-wave generation in Phillips theory. This hypothesis assumes that turbulence structures are convected by the mean flow without significant temporal evolution, allowing us to treat the pressure fluctuations as coherent over the relevant time scales. \citet{choi1990space} evaluated the Taylor’s frozen turbulence hypothesis, demonstrating its overall effectiveness, particularly for large turbulence structures. This convective nature of turbulence pressure fluctuations has been extensively studied in the turbulence literature (e.g., Choi \& Moin~\citeyear{choi1990space}; Luhar, Sharma \& MeKoen~\citeyear{luhar2014structure}; Yang \& Yang~\citeyear{yang2022wavenumber}).}
Given the initial conditions of $\hat\eta(\bm k,t=0)=0$ and $\partial_t\hat\eta(\bm k ,t=0)=0$, the solution to $\hat\eta$ is calculated as
	\begin{align}
		\hat\eta(\bm k,t)=\frac{k \hat{p}_0\left[-\mathrm{i}\sin(\Lambda t)\bm k\cdot\bm U+\Lambda\cos(\Lambda t)-\Lambda\exp(-\mathrm{i}\bm k\cdot\bm U t)
			\right]}{\Lambda[-(\bm k\cdot \bm U)^2+\Lambda^2]},
	\end{align}
where $k=|\bm k|$ is the magnitude of the wavenumber vector $\bm k$.
The wave energy component $|\hat\eta|^2$ has the following explicit expression:
	\begin{align}
		|\hat\eta|^2(\bm k,t)=&\frac{\Phi_p k^2}{\Lambda^2[(\bm k\cdot\bm U)^2-\Lambda^2]^2}\left[(\bm k\cdot\bm U)^2+\Lambda^2-2\Lambda^2\cos(\Lambda t)\cos(\bm k\cdot\bm U t)\right.\notag\\
		&\left.-2\Lambda\sin(\Lambda t)\bm k\cdot\bm U\sin(\bm k\cdot\bm U t)-((\bm k\cdot\bm U)^2-\Lambda^2)\cos^2(\Lambda t)\right],\label{eq:eta2_1}
	\end{align}
where $\Phi_p=|\hat{p}_0(\bm k)|^2$ is the energy spectrum of air pressure fluctuations at the air--water interface.
The surface elevation variance is obtained from the integral of $|\hat\eta|^2(\bm k,t)$ over the wavenumber space as follows:
	\begin{align}
		\langle\eta^2\rangle(t)= \iint |\hat\eta|^2(\bm k,t) \mathrm{d}\bm k.\label{eq:int_eta2_general}
	\end{align}
The goal of the present study is to understand the temporal behaviours of the wave energy, i.e., the surface elevation variance $\langle\eta^2\rangle (t)$. 
Next, we review the original Phillips approach for analysing wave energy growth, discuss its limitations, and introduce our new approach based on complex analyses.
 
\subsection{Review of the Phillips approach and its limitations}\label{sec2.2}
\begin{figure}
	\centering
         \captionsetup{width=1\linewidth, justification=justified}
        \includegraphics[width=0.99\textwidth,trim={0 3cm 0 3cm},clip]{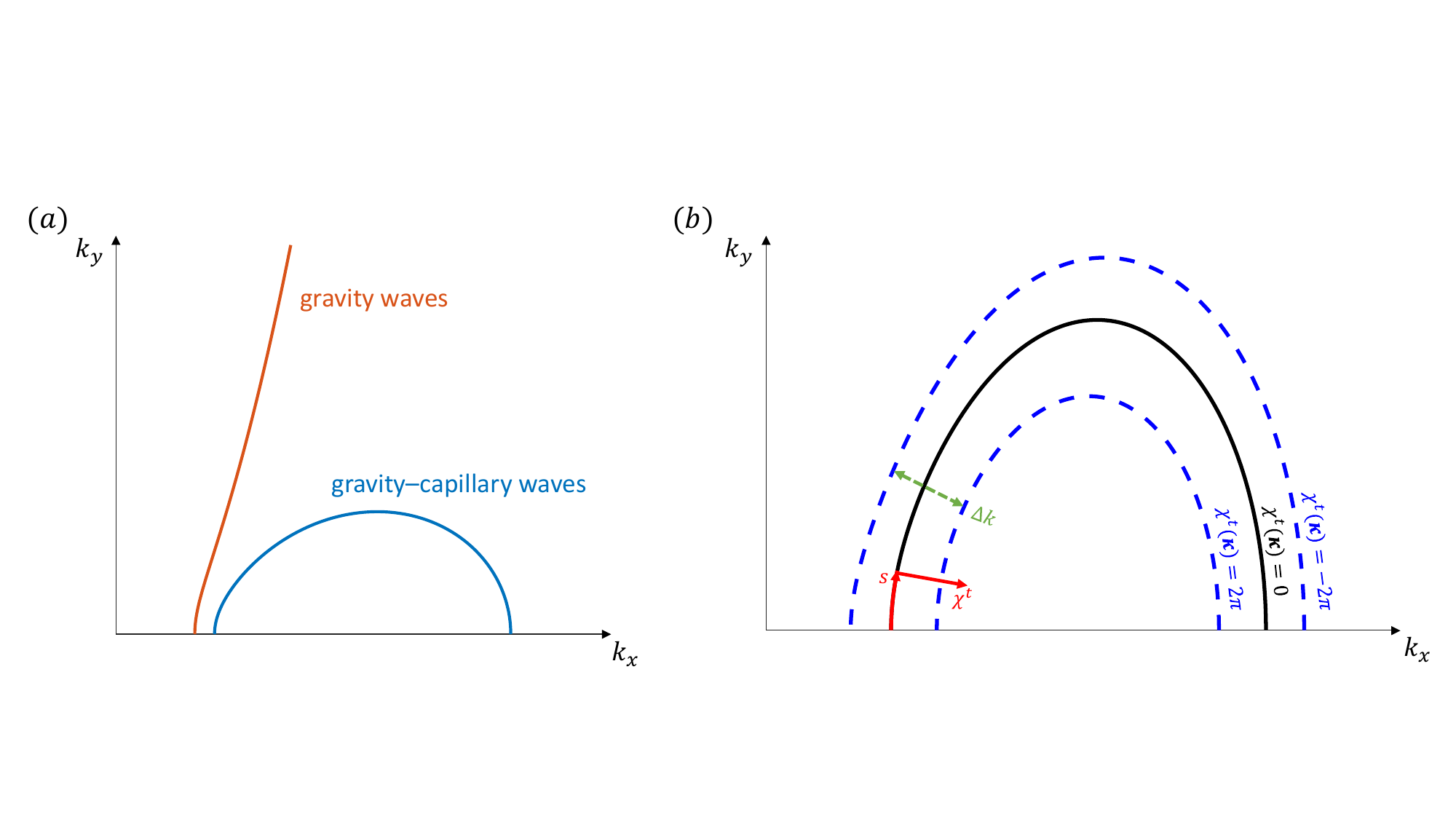}
	\caption{Panel (a) shows the resonance curves for gravity waves and gravity--capillary waves to illustrate their difference. Panel (b) illustrates the \citet{phillips1957generation} approach for calculating $\langle\eta^2\rangle$ based on the $(\chi^t,s)$ coordinate system in the wavenumber space. The content is a rework of figure 2 in \citet{phillips1957generation}.
    } \label{phillips_fig}
\end{figure}

In this section, we present a brief overview of the Phillips approach for analysing the evolution of wave energy (i.e., equation~\eqref{eq:int_eta2_general}). 
The wave energy component increases quadratically over time when it is located on the resonance curve in the spectral space, defined as $\bm k\cdot\bm U-\Lambda(\bm k)=0$. However, the wave growth rate decreases when the wavenumber deviates from this resonance curve.
To investigate the temporal behaviour of the wave energy, \citet[][]{phillips1957generation} utilised a new orthogonal curvilinear coordinate system that is associated with the resonance curve $\bm k\cdot\bm U-\Lambda(\bm k)=0$. 
Figure~\ref{phillips_fig}($a$) shows the resonance curves for gravity waves and gravity--capillary waves. 
Phillips introduced a new local orthogonal coordinate system $(\chi^t,s)$, where $\chi^t=\left(\bm k\cdot\bm U-\Lambda(\bm k)\right)t$ represents the nondimensional distance to the resonance curve and $s$ denotes the distance along the curve $\chi^t=0$ from the intersection of this curve with the $k_x$-axis. 
The coordinate system $(\chi^t,s)$ is illustrated using red arrows in figure~\ref{phillips_fig}($b$). 
The integral of the wave energy components over the two-dimensional wavenumber space can hence be calculated in this new $(\chi^t,s)$ coordinate system by incorporating Jacobian matrices. 
\citet{phillips1957generation} aimed to demonstrate that the integral of the wave energy components along $\chi^t$ between $\chi^t=-2\pi$ and $\chi^t=2\pi$ is proportional to time, resulting in a linear increase in the wave energy.  However,  this approach has several limitations and lacks rigour in the following aspects.
 \begin{itemize}
     \item First, the $(\chi^t,s)$ coordinate system is not explicitly defined; thus deriving the Jacobian matrix that connects the $(k_x,k_y)$ and $(\chi^t,s)$ coordinate systems at any arbitrary location is a nontrivial task. 
     \item Second, in Phillips' analyses, the Jacobian matrix was approximated using its value on the resonance curve ($\chi^t=0$) for various $\chi^t$ and $s$ values. Phillips~(\citeyear{phillips1957generation}) refers to this approximation as `simple.' This approximation did not lead to a quantitative result.
     \item Third, the integration along $\chi^t$ was numerically restricted to the interval $(-2\pi,2\pi)$, as depicted using the green dashed arrow in figure~\ref{phillips_fig}($b$). With this constraint, the contributions of the wave energy components outside this range to the total wave energy were neglected.  In other words, the calculations were not performed across the entire wavenumber space.
\end{itemize}

Hence, \citet{phillips1957generation} only pointed out that the integration of the wave energy components along $\chi^t$ is a linear function of time under the above approximations, but he did not obtain a quantification of the temporal behaviour of the wave energy with rigorous mathematical proof. In his original paper, Phillips described his method as `a simple and useful approximation to the integral.'
The fundamental reason why the Phillips approach cannot yield more rigorous results lies in the choice of the resonance curve-based orthogonal curvilinear coordinate system $(\chi^t,s)$.
However, we find that this curvilinear coordinate system is not necessary for wave evolution analyses.
In the present study, we show that the integral of the wave energy components can be directly analysed in a Cartesian coordinate system $(k_x, k_y)$ using a complex analysis method, as described in \S~\ref{sec2.3}.
Based on this new mathematical formulation, we develop a concrete theoretical framework for the wind-wave generation mechanism in the initial stage for gravity waves and gravity--capillary waves, as discussed in \S~\ref{sec3} and \S~\ref{sec4}, respectively, with the flow physics elucidated.

\subsection{Our new approach}\label{sec2.3}
We find that an analysis performed directly using the Cartesian wavenumber coordinates $(k_x,k_y)$ is feasible for analytically obtaining the linear increase in the leading-order term of the upper-bound solution of the surface elevation variance over time. 
This approach differs from that of \citet{phillips1957generation}. 
We present the main mathematical framework and results of the present study in this section.  Extensive mathematical analyses and investigations into the flow physics are provided in \S\S~\ref{sec3} and \ref{sec4}.

We evaluate the temporal increase in the surface elevation variance. \Cref{eq:int_eta2_general}, which describes the surface elevation variance $\langle\eta^2\rangle$, can be calculated by integrating first over variable $k_x$ and then over $k_y$.
Let function $E(k_y,t)$ be the integral of the wave energy component $|\hat\eta|^2$ along $k_x=(-\infty,\infty)$, i.e.,
	\begin{align}
		E(k_y,t)=\int_{-\infty}^{\infty}  |\hat\eta|^2(k_x,k_y,t) \mathrm{d}k_x.\label{eq:def_E}
	\end{align}
Therefore, the surface elevation variance $\langle\eta^2\rangle$ can be expressed as
	\begin{align}
		\langle\eta^2\rangle = \int_{-\infty}^{\infty} E(k_y,t)\mathrm{d} k_y.\label{eq:eta2_E}
	\end{align}\\
\noindent In the present study, we prove that for any $k_y\in\mathbb{R}$, the leading-order term of the upper-bound solution of $E(k_y,t)$ is a linear function of time~$t$.
	
We rewrite the wave energy component $|\hat\eta|^2(k_x, k_y, t)$ in~\eqref{eq:eta2_1} as 
	\begin{align}
		|\hat\eta|^2(k_x,k_y,t)=&\frac{\Phi_p (k_x^2+k_y^2)}{\Lambda^2(B^2-\Lambda^2)^2}
		\left\{\frac{1}{2}B^2+\frac{3}{2}\Lambda^2-\Lambda(\Lambda+B)\cos[(\Lambda-B)t]\right.\notag\\
		&\left.-\Lambda(\Lambda-B)\cos[(\Lambda+B)t]-\frac{1}{2}(B^2-\Lambda^2)\cos(2\Lambda t)\right\},\label{eq:def_H}
	\end{align}
where $B=\bm k\cdot \bm U=k_xU$.
Because the pressure spectrum $\Phi_p$ is generally not an analytic function, we use the following inequality to move the term $\Phi_p$ outside of the integral in \eqref{eq:def_E}. Combining \eqref{eq:def_E} and \eqref{eq:def_H}, we obtain the following estimation:
	\begin{align}
		E(k_y,t)\leq \Vert \Phi_p (k_x^2+k_y^2)\Lambda^{-3} \Vert_{L^\infty_{x}}\int_{-\infty}^{\infty} F(k_x, k_y,t) \mathrm{d}k_x,\label{eq:est_E}
	\end{align}
    where $F(k_x,k_y,t)$ is constructed as
	\begin{align}
		F(k_x,k_y,t)=&\frac{\Lambda}{(B^2-\Lambda^2)^2}
		\left\{\frac{1}{2}B^2+\frac{3}{2}\Lambda^2-\Lambda(\Lambda+B)\cos[(B-\Lambda)t]\right.\notag\\
		&\left.-\Lambda(\Lambda-B)\cos[(\Lambda+B)t]-\frac{1}{2}(B^2-\Lambda^2)\cos(2\Lambda t)\right\}.\label{eq:def_F}
	\end{align}
 {Here, the notation $\Vert\cdot\Vert_{L_x^\infty}$ denotes the $L^\infty$-norm with respect to $x$. For a general function $f(x,y)$, we have
\begin{align}
    \Vert f \Vert_{L_x^\infty}&=\max_{x\in \mathbb{R}} |f(x,y)|.\label{eq:def_Lx_inf_norm}
\end{align}}
The inequality \eqref{eq:est_E} always holds due to the positivity of function $|\hat\eta|^2(k_x,k_y,t)$.
The pressure spectrum $\Phi_p$ can be considered a function that decays quickly when the wavenumber $\bm k$ approaches zero or infinity in the wavenumber space. Therefore, the $L^\infty$ norm of $\Phi_p (k_x^2+k_y^2)\Lambda^{-3}$ has a finite value. 
We choose a specific expression for $F(k_x,k_y,t)$ to ensure that the improper integral of $F(k_x,k_y,t)$ with respect to $k_x$ on the right-hand side of \eqref{eq:est_E} also has a finite value.
For gravity waves, $\Lambda$ and $B$ behave as $\Lambda\sim |\bm k|^{1/2}$ and $B\sim k_x$, respectively.  
For gravity--capillary waves, $\Lambda$ and $B$ behave as $\Lambda\sim |\bm k|^{3/2}$ and $B\sim k_x$, respectively, when $k_x$ approaches infinity (i.e. $k_x\rightarrow\infty$). 
Therefore, for any wavenumber $k_y$, the function $F$ behaves as $k_x^{-3/2}$ when $k_x\rightarrow\infty$ for both gravity waves and gravity--capillary waves. The $-3/2$ power-law decay ensures that the integral is finite. When $k_y=0$, $F$ behaves as $F\sim k_x^{-1/2}$. The integral is also finite when the endpoint of the integration interval is $k_x=0$. Therefore, the right-hand side of \eqref{eq:est_E} has a finite value.  

Because $F(k_x,k_y,t)$ is an even function with respect to $k_x$, we can limit our focus to the integral along the positive $x$-axis, i.e., the interval $(0,\infty)$.  Hence, we define function $G(k_y,t)$ as 
    \begin{align}
        G(k_y,t)=\int_0^\infty F(k_x,k_y,t) \mathrm{d}k_x. \label{eq:def G}
    \end{align}
The function $G(k_y,t)$ represents half of the integral of $F(k_x,k_y,t)$ with respect to $k_x$ over the range $(-\infty,\infty)$.

To understand the temporal evolution of the surface elevation variance, we should examine the right-hand side of \eqref{eq:est_E}, especially the function $G(k_y,t)$ in~\eqref{eq:def G}.
Next, we prove that for gravity waves, the function $G(k_y,t)$ is bounded by a linear function of time plus an oscillatory component with an amplitude that is restricted by a constant.
The linear increase in $G(k_y,t)$ suggests that the surface elevation variance $\langle\eta^2\rangle(t)$ increases linearly with time $t$. 
However, the linear increase in the surface elevation variance for gravity--capillary waves depends on whether the convection velocity of the air pressure fluctuations at the water surface exceeds a certain threshold $U_{\mathrm{min}}$.
Specifically, for gravity--capillary waves, when the convection velocity of the air pressure fluctuations at the water surface, $U$, exceeds $U_{\mathrm{min}}$, the function $G(k_y,t)$ is restricted by a linear function of time plus an oscillatory component with an amplitude that is limited by a constant.
Consequently, the surface elevation variance $\langle\eta^2\rangle$ exhibits linear growth over time.
However, when the magnitude of the convection velocity of the air pressure fluctuations, $U$, is less than $U_{\mathrm{min}}$, the function $G(k_y,t)$ is restricted by a constant that is independent of time $t$, which suggests that waves do not grow under such weak wind conditions.  The detailed proof is presented in \S~\ref{sec3} for gravity waves and \S~\ref{sec4} for gravity--capillary waves.

\section{Gravity waves}\label{sec3}
In this section, we examine the temporal evolution for the surface elevation variance of gravity waves, i.e., we do not consider the effect of surface tension.
We first demonstrate that for any $k_y\in \mathbb{R}$, there exists a corresponding $k_x$ such that the point $(k_x,k_y)$ lies on the resonance curve.
Next, we employ the residue theorem to prove that $G(k_y,t)$ is equal to $M(k_y,t) + N(k_y,t)$. Here, $M(k_y,t)$ is a linear function of time $t$, and $N(k_y,t)$ is an oscillatory function over time $t$ with an amplitude that is bounded by a constant. Finally, we show that the surface elevation variance $\langle\eta^2\rangle$ asymptotically increases linearly with time in the initial stage of Phillips theory.

\subsection{Properties of the resonance curve for gravity waves}
The dispersion relation of gravity surface waves relates the angular frequency $\Lambda$ of the waves to their wavenumber $\bm k$ and is given by $\Lambda =\sqrt{g|\bm k|}$, where $|\bm k|=\sqrt{k_x^2+k_y^2}$ and $g$ represents gravitational acceleration.
The denominator in \eqref{eq:def_F} is singular when the following resonance condition is satisfied: $B-\Lambda=0$. 
For a chosen $k_y\in\mathbb{R}$, the value of $k_x$ such that the point $(k_x,k_y)$ is located on the resonance curve $\chi=0$ in the wavenumber space must satisfy the following condition:
	\begin{align}
		k_x=\pm \frac{g}{U^2}\sqrt{\frac{1}{2}\left(1+\sqrt{1+\frac{4k_y^2U^4}{g^2}}\right)}.
	\end{align}
This implies that for any $k_y$, there exists a positive value of $k_x$ that satisfies the resonance condition $\Lambda - \bm k \cdot U = 0$.
In the first quadrant of the wavenumber space, the resonance curve $\chi$ begins at $(g/U^2,0)$ on the $x$-axis and extends to infinity, as depicted in figure~\ref{phillips_fig}($a$).

\subsection{Calculation of $G(k_y,t)$}	   
Next, we calculate the function $G(k_y,t)$ defined in \eqref{eq:def G}.
The key aspect of our approach is that we employ the residue theorem to calculate the integral in the definition of $G(k_y,t)$, as shown in \eqref{eq:def G}.
First, we perform an analytical continuation to convert the function $F(k_x,k_y,t)$ in the integrand on the right-hand side of \eqref{eq:def G} into a complex function with good properties. We calculate the leading-order term using the residue theorem, resulting in a linear function of time. The oscillatory term is obtained based on complex analyses.

\subsubsection{Analytical continuation of $F$ in the complex plane}
\begin{figure}
	\centering
\includegraphics[width=0.45\textwidth]{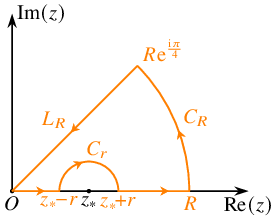}
	\caption{Illustration of the indented closed loop for integration, as outlined in~\eqref{eq:loop1}.\label{fig: loop gravity waves}}
\end{figure}

In this subsection, we present a novel approach to evaluate the integral of $F$ using complex analyses. 
We transform the real-valued function $F$ into a complex-valued function in the complex $z$-plane. Hereafter,  $z$ represents a complex number. 
We examine a related complex function $F_1(z,k_y,t)$, which is obtained by analytically continuing the function $F(k_x,k_y,t)$ into the complex $z$-plane as follows:
	\begin{align}
		F_1(z,k_y,t)=&\frac{\Lambda}{(B^2-\Lambda^2)^2}
		\left\{\frac{1}{2}B^2+\frac{3}{2}\Lambda^2-\Lambda(\Lambda+B)\exp[\mathrm{i}(B-\Lambda)t]\right.\notag\\
		&\left.-\Lambda(\Lambda-B)\exp[\mathrm{i}(\Lambda+B)t]-\frac{1}{2}(B^2-\Lambda^2)\exp(2\mathrm{i}\Lambda t)\right\}.\label{eq:F1complex}
	\end{align}
Here, $B=Uz$ and $\Lambda=\sqrt{g\sqrt{z^2+k_y^2}}$, as we have substituted the variable $k_x$ with the complex-valued variable $z$.
It is evident that $F(k_x,k_y,t)$ is equal to the real part of $F_1(k_x,k_y,t)$ for any real values of $k_x$, $k_y$, and $t$.
The function $F_1$ has a pole at $z=z_*$ on the positive real axis, where $B - \Lambda = 0$:
    \begin{align}
	z_*  =\frac{g}{U^2}\sqrt{\frac{1}{2}\left(1+\sqrt{1+\frac{4k_y^2U^4}{g^2}}\right)}.
    \end{align}
We consider a closed loop $\Gamma$ in the complex plane, as depicted in figure~\ref{fig: loop gravity waves}, which is defined as $\Gamma=[0, z_*-r]+C_r+[z_*+r, R]+C_R+L_R$.
Here, $C_r( {\theta})=z_*+re^{\mathrm{i}(\pi-\theta)}$, with $0\leq \theta \leq \pi$, is a clockwise circular path around $z_*$ with a radius of $r$; $C_R(\theta)=Re^{\mathrm{i}\theta}$, with $0\leq \theta \leq \pi/4$, is an anticlockwise circular path around 0 with a radius of $R$; 
and $L_R(\theta)=Re^{\mathrm{i}\pi/4}(1-\theta)$, with $0\leq\theta\leq 1$, is a straight path pointing towards the origin.
We choose the angle between the real axis and the ray $-L_R$ to be $\pi/4$. 
However, the choice of this angle is arbitrary, and the real part of the integral of the function $F_1$ along $L_R$ is not dependent on this angle.
This is ensured by the fact that the contribution of intergral along $C_R$ vanishes as $R\rightarrow\infty$ regardless the angle between $L_R$ and the $x$-axis, as further discussed in~\S\,\ref{sec:3.2.2}. The independence of the angle can also be confirmed through numerical integration.
Because the function $F_1$ has no singularities inside the closed loop $\Gamma$, the integral of $F_1$ with respect to $z$ along the contour $\Gamma$ is equal to zero. This can be written as
\begin{align}
	\left(\int_0^{z_*-r}+\int_{C_r}+\int_{z_*+r}^R+\int_{C_R}+\int_{L_R}\right)F_1(z,k_y,t)\mathrm{d}z=0.\label{eq:loop1}
\end{align}
Therefore, the integration of $F_1$ along the real positive axis can be calculated as
\begin{align}
	\int_0^{\infty}F_1(k_x,k_y,t) \mathrm{d}k_x=&\lim_{R\rightarrow\infty}\int_{0}^{R}F_1(k_x,k_y,t) \mathrm{d}k_x\notag\\
	=&\lim_{R\rightarrow\infty}\lim_{r\rightarrow0}\mathrm{Re}\left[\left(\int_0^{z_*-r}+\int_{z_*+r}^R\right)F_1(z,k_y,t)\mathrm{d}z\right]\notag\\
	=&\lim_{R\rightarrow\infty}\lim_{r\rightarrow0}\mathrm{Re}\left[\left(-\int_{C_r}-\int_{C_R}-\int_{L_r}\right)F_1(z,k_y,t)\mathrm{d}z\right]\notag\\
	=&-\lim_{R\rightarrow\infty}\mathrm{Re}\left[\int_{C_R}F_1(z,k_y,t) \mathrm{d}z\right]
	-\lim_{r\rightarrow0}\mathrm{Re}\left[\int_{C_{r}}F_1(z,k_y,t) \mathrm{d}z\right]\notag\\
	&-\lim_{R\rightarrow\infty}\mathrm{Re}\left[\int_{C_{L_R}}F_1(z,k_y,t) \mathrm{d}z\right].\label{eq:int_F1_expression}
\end{align}

\subsubsection{Calculation of the integral of $F_1$ along the path $C_R$}\label{sec:3.2.2}
We evaluate $F_1$ along the path $C_R$. Upon examining the expression of $F_1$ in equation (\ref{eq:F1complex}), we observe that the essential properties of $F_1$ are governed by three exponential functions, namely, $\exp[\mathrm{i}(B-\Lambda)t]$, $\exp[\mathrm{i}(B+\Lambda)t]$, and $\exp(2\mathrm{i}\Lambda t)$.
Assuming $z = R\exp(\mathrm{i}\theta)$ with $0\leq\theta\leq \pi/4$, for sufficiently large values of $R$, we have the following asymptotic expansion:
\begin{align}
	B-\Lambda&=UR\mathrm{e}^{\mathrm{i}\theta}-\sqrt{g}\left(
	R^2 \mathrm{e}^{2\mathrm{i}\theta}+k_y^2\right)^{\frac{1
		}{4}}\notag\\
	&=UR\mathrm{e}^{\mathrm{i}\theta}-\sqrt{gR} \mathrm{e}^{\frac{\mathrm{i}\theta}{2}}+O\left(R^{-\frac{3}{2}}\right).\notag
\end{align}
Therefore, for sufficiently large values of  $R$, we obtain the following expression for $B-\Lambda$:
\begin{align}
	B-\Lambda=\alpha_1+\mathrm{i}\beta_1,
\end{align}
where $\alpha_1$ and $\beta_1$ are real positive functions of $R$. 
The leading-order approximation shows that $\alpha_1\approx UR\cos(\theta)$ and $\beta_1\approx UR\sin(\theta)$ for sufficiently large values of $R$. Hence, the term $\exp[\mathrm{i}(B-\Lambda)t]$ becomes
\begin{align}
	\exp[\mathrm{i}(B-\Lambda)t]=\exp(-\beta_1 t+\mathrm{i}\alpha_1 t).
\end{align}
This analysis demonstrates that the real part of $\exp[\mathrm{i}(B-\Lambda)t]$ is a bounded function and it decays as $R$ or $t$ approaches infinity.

Using a similar method, we next study the properties of the function  {$\exp[\mathrm{i}(B+\Lambda)t]$}.  Because $0\leq\theta\leq\pi/4$, we have the following expansion for $B+\Lambda$:
\begin{align}
	B+\Lambda=\alpha_2+\mathrm{i}\beta_2,
\end{align}
where $\alpha_2$ and $\beta_2$ are both real positive functions of $R$, and the leading-order approximation shows that $\alpha_1\approx UR\cos(\theta)$ and $\beta_1\approx UR\sin(\theta)$ for sufficiently large $R$.
The term $\exp[\mathrm{i}(B+\Lambda)t]$ becomes
\begin{align}
	\exp[\mathrm{i}(B+\Lambda)t]=\exp(-\beta_2 t-\mathrm{i}\alpha_2 t).
\end{align}
This result indicates that the real part of $\exp[\mathrm{i}(B+\Lambda)t]$ is a bounded function and it decays as $R$ or $t$ approaches infinity.

For the term $\exp(2\mathrm{i}\Lambda t)$, we calculate the following asymptotic expansion for a sufficiently large value of $R$:
\begin{align}
	2\mathrm{i}\Lambda=\alpha_3+\mathrm{i}\beta_3,
\end{align}
where $\alpha_3$ and $\beta_3$ are positive functions of $R$ that satisfy $\alpha_3\approx2\sqrt{gR}\cos(\theta/2)$ and $\beta_3\approx2\sqrt{gR}\sin(\theta/2)$, respectively, for sufficiently large values of $R$. 
Hence, the term $\exp(2\mathrm{i}\Lambda t)$ becomes
\begin{align}
	\exp(2\mathrm{i}\Lambda t)=\exp(-\beta_3 t-\mathrm{i}\alpha_3 t),
\end{align}
of which the real part is also a bounded function of the variables $R$ and $t$.

The above analyses suggest that, along the curve $C_R$,
\begin{align}
	|F_1|<\frac{C}{R^2},
\end{align}
where $C$ is a function of $k_y$ and $t$ and does not depend on $R$.  Therefore, we obtain the following estimation:
\begin{align}
0\leq\left|\lim_{R\rightarrow\infty}\int_{C_R}F_1 \mathrm{d}z\right|
	\leq \lim_{R\rightarrow\infty}\int_{C_R}\left|F_1(z)\right| \mathrm{d}z
	\leq\lim_{R\rightarrow\infty} \int_{C_R} \frac{C}{R^2}\mathrm{d}z=0.
\end{align}
We thus conclude that
\begin{align}
	\lim_{R\rightarrow\infty}\int_{C_R}F_1(z) \mathrm{d}z=0.\label{eq:gw_term1}
\end{align}

\subsubsection{Calculation of the integral of $F_1$ along the path $C_r$}\label{sec:3.2.3}
The integral of $F_1$ along the semicircular path $C_r$ can be calculated using the residue theorem, from which we have
\begin{align}
	\lim_{r\rightarrow0}\int_{C_{r}}F_1(z) \mathrm{d}z&=-\pi\mathrm{i}\mathrm{Res}(F_1,z_{*}).
\end{align}
Here, the notation $\mathrm{Res}(F_1,z_{*})$ denotes the residue of the function $F_1$ at the point $z=z_*$.
The function $F_1$ has a first-order pole at $z = z_*$. As a result, we have
\begin{align}
	\mathrm{Res}(F_1,z_*)=\lim_{z\rightarrow z_*} F_1(z)(z-z_*).\label{eq:def_Res_F1}
\end{align}
To determine the residue of the function $F_1$ in the above equation, we employ a perturbation method to calculate the limit near $z=z_*$.
We first perform a Taylor expansion of the term $B-\Lambda$ around $z=z_*$ and obtain
\begin{align}
	B-\Lambda=\frac{\left(z_*^2+2k_y^2\right)U}{2z_*^2+2k_y^2}(z-z_*)+O\left((z-z_*)^2\right).
\end{align}
The same procedure is applied to both the denominator and numerator of the function $F_1(z,k_y,t)$ near $z=z_*$. For simplicity, we define a dimensionless local Froude number $\gamma_y$ based on the wavenumber $k_y$ and convection velocity of the pressure fluctuations $U$ as
\begin{align}
	\gamma_y=\frac{U^2 k_y}{g}.\label{eq:def_fr_loc}
\end{align}
Hence, we obtain that 
\begin{align}
	 {\lim_{r\rightarrow0}\int_{C_{r}}F_1(z,k_y,t) \mathrm{d}z=-\frac{\pi g\left(1+\sqrt{4\gamma_y^2+1}\right)^{\frac{3}{2}}}{2\sqrt{2}\sqrt{4\gamma_y^2+1}U^2}t}.\label{eq:linear_component_gravity waves}
\end{align} 
Equation \eqref{eq:linear_component_gravity waves} indicates that this component is linearly proportional to time $t$.

\subsubsection{Calculation of the integral of $F_1$ along the path $L_R$}
Next, we consider the integral of $F_1$ along path $L_R$, where the variable $z$ can be parameterised as a function of a real variable $r$, such that $z(r) = r\exp(\mathrm{i}\pi/4)$ with $0\leq r\leq R$.  Therefore, we have
\begin{align}
	\int_{L_R} F_1(z)\mathrm{d}z&=\exp\left(\frac{\mathrm{i}\pi}{4}\right)\int_R^0 F_1(z(r))\mathrm{d}r\notag\\
	&=-\frac{1+\mathrm{i}}{\sqrt{2}}\int_0^R\frac{K(r,t)}{J(r)}\mathrm{d}r.\label{eq:LR10}
\end{align}
In \eqref{eq:LR10}, the numerator $K(r,t)$ and denominator $J(r)$ are
\begin{align}
	K(r,t)=&\Lambda_0(r)\left\{\frac{1}{2}B_0(r)^2+\frac{3}{2}\Lambda_0(r)^2-\Lambda_0(r)[\Lambda_0(r)+B_0(r)]\exp[\mathrm{i}(B_0(r)-\Lambda_0(r))t]\right.\notag\\
	&-\Lambda_0(r)[\Lambda_0(r)-B_0(r)]\exp[\mathrm{i}(B_0(r)+\Lambda_0(r))t]\notag\\
	&\left.-\frac{1}{2}\left[B_0(r)^2-\Lambda_0(r)^2\right]\exp[2\mathrm{i}\Lambda_0(r) t]\right\}
 \end{align}
and
\begin{align}
	J(r)=&\left[B_0(r)^2-\Lambda_0(r)^2\right]^2,\label{eq:expr_Jr-0}
\end{align}
respectively.
Here, the complex-valued variables $B_0(r)$ and $\Lambda_0(r)$ are functions of the real-valued number $r$. They can be written as
\begin{align}
	B_0(r)&=\frac{1+\mathrm{i}}{\sqrt{2}}Ur,\\
	\Lambda_0(r)&=\sqrt{g\sqrt{\mathrm{i}r^2+k_y^2}}.
\end{align}
We aim to examine the time evolution behaviour of $F_1(z)$ for any complex variable $z$ located along the path $L_R$.
For large values of $k_y$, $\mathrm{Im}(B_0(r) - \Lambda_0(r)) > 0$ always holds.
We can split the complex-valued term $B_0(r) - \Lambda_0(r)$ into a real part $\alpha_4(r)$ and an imaginary part $\beta_4(r)$:
\begin{align}
	B_0(r)-\Lambda_0(r)=\alpha_4(r)+i\beta_4(r).
\end{align}
Here, $\alpha_4(r)$ and $\beta_4(r)$ are real functions and $\beta_4(r)\ge 0$.  Hence we have
\begin{align}
	\exp[\mathrm{i}(B_0(r)-\Lambda_0(r))t]=\exp[\mathrm{i}\alpha_4(r)t]\exp[-\beta_4(r)t]. \label{eq:split1}
\end{align}
We observe that the magnitude of $\exp\left(\mathrm{i}(B_0(r) - \Lambda_0(r))t\right)$ is a bounded function of time $t$ for all complex variables $z(r)$ located along the path $L_R$:
\begin{align}
	\left|\exp\left[\mathrm{i}(B(r)-\Lambda(r))t\right]\right|\leq 1. \label{eq:est B-Lambda}
\end{align}
Similarly, we can obtain that
\begin{align}
		\left|\exp[\mathrm{i}(B(r)+\Lambda(r))t]\right|&\leq 1,\\
		\left|\exp[\mathrm{i}(2\Lambda(r))t]\right|&\leq 1.
\end{align}
Therefore, by using the triangle inequality, the modulus of the complex-valued function $K(r)$ is bounded by
\begin{align}
	|K(r,t)|\leq |B_0(r)|^2|\Lambda_0(r)|+4|\Lambda_0(r)|^3+2|B_0(r)\Lambda_0^2(r)|.\label{eq:kr}
\end{align}
According to \eqref{eq:split1}, $K(r,t)$ has the following asymptotic behaviour for large values of $t$:
\begin{align}
	\lim_{t\rightarrow\infty}K(r,t)=\frac{1}{2}B_0^2(r) {\Lambda_0(r)}+\frac{3}{2}\Lambda_0^3(r).
\end{align}
We note that the right-hand side of \eqref{eq:kr} is independent of time $t$.  The denominator $J(r)$ on the right-hand side of \eqref{eq:LR10} is positive for all $z$ on $L_R$, indicating that equation (\ref{eq:LR10}) does not have singularities along the path $L_R$. Therefore, from equation (\ref{eq:LR10}), we can estimate its modulus as follows:
\begin{align}
	\left|\int_{L_R}F_1(z)\mathrm{d}z\right|
	&\leq\int_0^\infty\frac{|B_0(r)|^2|\Lambda_0(r)|+4|\Lambda_0(r)|^3+2|B_0(r)\Lambda_0^2(r)|}{|B_0(r)^2-\Lambda_0(r)^2|^2}\mathrm{d}r.\label{eq:LR2}
\end{align}
The integrand on the right-hand side of equation \eqref{eq:LR2} has an asymptotic behaviour of $r^{-1.5}$ as $r$ approaches infinity. Consequently, the integration over the path $L_R$ yields a finite value that depends on the parameters $U$, $g$, and the local spanwise wavenumber $k_y$. 
When $k_y$ is small, the positivity of $\beta_4(r)$ cannot hold true for all values of $r>0$, although the decaying property of $B_0(r)-\Lambda_0(r)$ ensures that $\beta_4(r)$ is positive for large values of $r$. As a result, under such circumstances, the pointwise estimation of \eqref{eq:est B-Lambda} is not valid with respect to $r$ due to the potential exponential growth of time $t$. However, the function $F(k_x,k_y,t)$ exhibits at most quadratic growth over time, and thus the exponential growth term should cancel out during the integration of $F_1$ along the path $L_R$. In numerical examples in the next subsection, we confirm that the integral of $F_1$ along the path $L_R$ shows oscillatory behaviour with a decaying amplitude.

\subsubsection{Properties of the function $G$}
In this subsection, we summarise the calculations of $G(k_y,t)$ as defined in~\eqref{eq:def G}.
Recall that integrating the function $F_1$ with respect to $z$ along the positive real axis yields the same result as the function $G(k_y,t)$ (see~\cref{eq:int_F1_expression,eq:gw_term1,eq:linear_component_gravity waves,eq:LR10}).  
Then, we obtain the following expression for the function $G(k_y,t)$:
\begin{align}
	G(k_y,t)=M(k_y,t)+N(k_y,t),\label{eq:G M N}
\end{align}
where the functions $M(k_y,t)$ and $N(k_y,t)$ are
\begin{align}
    M(k_y,t)&= {\frac{\pi g\left(1+\sqrt{4\gamma_y^2+1}\right)^{\frac{3}{2}}}{2\sqrt{2}\sqrt{4\gamma_y^2+1}U^2}t},\\
    N(k_y,t)&=\mathrm{Re}\left[\int_0^\infty\frac{1+\mathrm{i}}{\sqrt{2}}\frac{K(r,t)}
	{ {J}(r)}
	\mathrm{d}r\right].
\end{align}
{Here, $\gamma_y$ is the local Froude number defined in \eqref{eq:def_fr_loc}, $N(k_y,t)$ is the component that deviates from linear growth, and $J(r)$ is defined in~\eqref{eq:expr_Jr-0}.}
The function $N(k_y,t)$ oscillates with time $t$, but its amplitude is bounded by a constant, as defined on the right-hand side of \eqref{eq:LR2}. 
 {Let $q(k_y)$ be the linear growth rate of $M(k_y,t)$:
\begin{align}
q(k_y)=\frac{\pi g\left(1+\sqrt{4\gamma_y^2+1}\right)^{\frac{3}{2}}}{2\sqrt{2}\sqrt{4\gamma_y^2+1}U^2}.\label{eq:def:q_ky}
\end{align}
Because $\gamma_y$ scales as $\gamma_y\sim U^2$, we obtain that the linear growth rate $q(k_y)$ has the following approximations for small $\gamma_y$ and large $\gamma_y$:
\begin{align}
    q(k_y)\approx\begin{cases}
         \frac{\pi g}{2\sqrt{2}U^2}, &\gamma_y\ll 1\\       \frac{\pi\sqrt{gk_y}}{2U}, &\gamma_y\gg 1\label{eq:q_ky_scaling}
    \end{cases}\
\end{align}
which shows that $q(k_y)$ scales as $q(k_y) \sim U^{-2}$ for small $\gamma_y$ and $q(k_y) \sim U^{-1}$ for large $\gamma_y$.}
Additionally, as time $t$ approaches infinity, we have the following limit result:
\begin{align}
	\lim_{t\rightarrow\infty}N(k_y,t)=\mathrm{Re}\left[
	\int_0^\infty \frac{(1+\mathrm{i})\sqrt{g}\left(\mathrm{i}U^2r^2+3g\sqrt{\mathrm{i}r^2+k_y^2}\right)\left(\mathrm{i}r^2+k_y^2\right)^{\frac{1}{4}}}{4\left(\mathrm{i}U^2r^2-g\sqrt{\mathrm{i}r^2+k_y^2}\right)^2}\mathrm{d}r
	\right]\label{eq:N_limit}
\end{align}

 \begin{figure}
	\centering
         \captionsetup{width=1\linewidth, justification=justified}
        \includegraphics[width=1\textwidth,trim={1cm 0 1.2cm 0},clip]{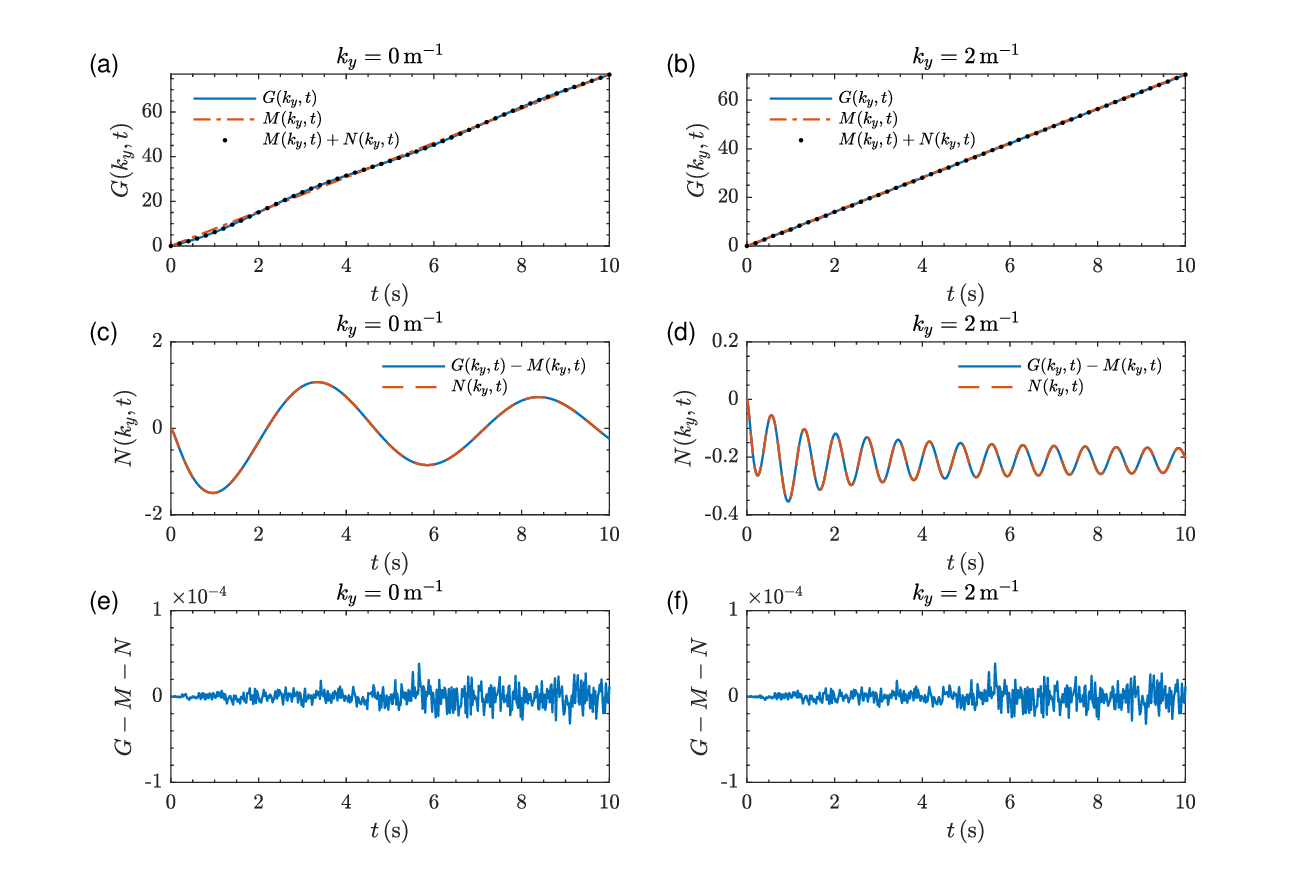}
	\caption{Numerical examples of the calculation of $G(k_y,t)$ for representative values of the wavenumber $k_y$, with results for $k_y=0\,\mathrm{m^{-1}}$ shown in the left column (panels a, c, and e) and results for $k_y=2\,\mathrm{m^{-1}}$ shown in the right column (panels b, d, and f). We set $g=9.81\,\mathrm{m\,s^{-2}}$ and $U=2\,\mathrm{m\,s^{-1}}$. Panels (a) and (b): $G(k_y,t)$, with the blue lines indicating the results obtained from numerical integration, the red lines indicating the linear growth component, $M(k_y,t)$, and the black dots representing the sum of the linear growth component $M(k_y,t)$ and the oscillatory component $N(k_y,t)$. Panels (c) and (d): the red lines show the oscillatory component $N(k_y,t)$, and the blue lines show $G(k_y,t)-M(k_y,t)$. Panels (e) and (f): numerical errors in the calculations, with the blue lines representing $G(k_y,t)-M(k_y,t)-N(k_y,t)$.
    } \label{compare_gravity_waves}
\end{figure}

 We present numerical examples for calculating the function $G(k_y,t)$ for gravity waves, as shown in figure \ref{compare_gravity_waves}. To verify \eqref{eq:G M N}, we compute the function $G(k_y,t)$ by performing a numerical integration of \eqref{eq:def G} and compare the result to the sum of $M(k_y, t)$ and $N(k_y, t)$.
This comparison is illustrated in panels (a) and (b) of figure \ref{compare_gravity_waves}.
The results confirm that the function $G(k_y, t)$ predominantly exhibits linear growth over time and that the linear function $M(k_y, t)$ perfectly captures the leading-order behaviour of $G(k_y, t)$.  
The time evolution of the function $N(k_y,t)$, which is the oscillatory component of $G(k_y,t)$, is shown in panels (c) and (d) of figure \ref{compare_gravity_waves}.  As $N(k_y,t)$ oscillates, its amplitude gradually decreases over time. 
This phenomenon supports the conclusion that the function $N(k_y,t)$ converges to a finite limit value as time approaches infinity, as demonstrated by \eqref{eq:N_limit}.  
Panels (e) and (f) of figure \ref{compare_gravity_waves} show the numerical error for the integration of $G(k_y,t)$, calculated as $G(k_y,t)-M(k_y,t)-N(k_y,t)$. 
The difference between $G(k_y,t)$ and $M(k_y,t)+N(k_y,t)$ remains less than $10^{-4}$ while $G(k_y,t)$ ranges from $0$ to $60$, thereby validating the theoretical analyses.

\subsection{Overall increases in the surface elevation variance $\langle\eta^2\rangle$ over time}
In this subsection, we demonstrate the evolution of the surface elevation variance $\langle\eta^2\rangle$ based on equation \eqref{eq:eta2_E}. 
By combining \eqref{eq:est_E}, \eqref{eq:def G} and \eqref{eq:G M N},
we obtain the following estimation of the function $E(k_y,t)$:
\begin{align}
    E(k_y,t)\leq 2\Vert \Phi_p (k_x^2+k_y^2)\Lambda^{-3} \Vert_{L^\infty_{x}}[M(k_y,t)+N(k_y,t)].\label{eq:E2}
\end{align}

Hence, the surface elevation variance $\langle\eta^2\rangle$ can be calculated from \eqref{eq:eta2_E} as
\begin{align}
	\langle\eta^2\rangle
        \leq& 2\int_{-\infty}^\infty 
                \Vert \Phi_p(k_x^2+k_y^2)\Lambda^{-3}\Vert_{L_x^\infty}
                [M(k_y,t)+N(k_y,t)]
                \mathrm{d}k_y\notag\\
	\leq& \left\Vert\Phi_p(k_x^2+k_y^2)\Lambda^{-3}q(k_y)
                \right\Vert_{L_x^\infty L_y^1} t\notag\\
            &+2\int_{-\infty}^\infty 
                \Vert \Phi_p(k_x^2+k_y^2)\Lambda^{-3}\Vert_{L_x^\infty} N(k_y,t) 
                \mathrm{d}k_y,\label{eq:eta2_est_gravitywave}
\end{align}
where $\gamma_y$ and $q(\gamma_y)$ are defined in \eqref{eq:def_fr_loc} and \eqref{eq:def:q_ky}, respectively.
Here, the notation of $\Vert\cdot\Vert_{L_x^\infty L_y^1}$ denotes the $L^\infty$-norm with respect to $x$ and $L^1$-norm with respect to $y$. For a general function $f(x,y)$, we have
\begin{align}
     {\Vert f \Vert_{L_x^\infty L_y^1}=\int_{-\infty}^\infty\max_{x\in \mathbb{R}} |f(x,y)|\mathrm{d}y.}
\end{align}
\noindent Equation~\eqref{eq:eta2_est_gravitywave} explicitly demonstrates the upper bound of the temporal increase in the energy of gravity waves; that is, the leading-order term increases linearly over time, and there also exists an oscillatory term representing the fluctuations in the surface elevation variance.
 {The pressure fluctuation spectrum $\Phi_p$ is proportional to $u_\tau^4$, where $u_\tau$ is the air friction velocity~\citep{phillips1957generation}.  Considering the scaling of $q(k_y)$ with respect to $U$ (see \eqref{eq:q_ky_scaling}), we can presumably infer that the wave energy $\langle\eta^2\rangle$ scales as $\langle\eta^2\rangle\sim U^s$ with $s\in[2,3]$ during the initial stage of gravity wave growth.}

In conclusion, we calculate the surface elevation variance $\langle\eta^2\rangle$ for gravity waves forced by wind by integrating the wave energy component $|\hat\eta(\bm k)|^2$ in the wavenumber space.  We apply a complex analysis method to transform the integration along the $k_x$ axis to a singular integration within the complex $z$-plane and resolve the singularity using the residue theorem. Our results indicate that for gravity waves forced by wind, the leading-order term of the upper-bound solution of the surface elevation variance is a linear function of time, while the sub-leading term shows oscillatory behaviour.

\section{Gravity--capillary waves}\label{sec4}
In this section, we evaluate the temporal evolution of gravity--capillary waves when surface tension effects are considered.
Hence, the dispersion of gravity--capillary waves can be written as
\begin{align}
    \Lambda (\bm k)=\sqrt{g|\bm k|+(\sigma/\rho_w)|\bm k|^3}.\label{eq:def_lambda_gcwave}
\end{align}
Here, $\sigma$ denotes the surface tension.  
The additional cubic term $(\sigma/\rho_w)|\bm k|^3$ in \eqref{eq:def_lambda_gcwave} introduces a high-order nonlinearity that characterises the shape of the resonance curve, as demonstrated in the following subsections.
When considering surface tension, the phase velocity of the gravity--capillary wave becomes
\begin{align}
    c(\bm k)=\sqrt{\frac{g}{|\bm k|}+\frac{\sigma}{\rho_w}|\bm k|}.\label{eq:def:c_gcwave}
\end{align}
Equation~\eqref{eq:def:c_gcwave} shows that the wave phase velocity has a global minimum $c_{\mathrm{min}}$ at the critical wavenumber $k=k_0$:
\begin{align}
    k_0&=\sqrt{\frac{g\rho_w}{\sigma}},\\
    c_{\mathrm{min}}&=\sqrt{2\sqrt{\frac{g\sigma}{\rho_w}}}.
\end{align}
In the following subsections, we demonstrate the important effects of the critical wavenumber $k_0$ and the minimum phase velocity $c_{\mathrm{min}}$ on the shape of the resonance curve and the characterisation of the temporal increase in the wave energy.
 {Given the physical properties of $g=9.8\,\mathrm{m\,s^{-2}}$, $\rho_w=10^3\,\mathrm{kg\,m^{-3}}$ and $\sigma=7\times10^{-2}\,\mathrm{kg\,s^{-2}}$, the minimum phase velocity is $c_{\mathrm{min}}=0.23~\mathrm{m\,s^{-1}}$. \cite{perrard2019turbulent} showed that the convection velocity of air pressure fluctuations at the air--water interface $U$ lies in the range of $[0.5, 0.8]U_a$, where $U_a$ is the mean wind velocity for a boundary layer thickness of $3\,\mathrm{cm}$.  The convection velocity $U$ can also be related with the air friction velocity $u_\tau$ as $U\in [12, 16]u_\tau$~\citep{li2022principal}. Therefore, when the convection velocity is close to the critical value of $c_{\mathrm{min}}$, the corresponding wind velocity $U_a$ is approximately $U_a\in[0.29, 0.46]\,\mathrm{m \, s^{-1}}$, and the air friction velocity $u_\tau^a$ is approximately $u_\tau\in[0.014, 0.019]\,\mathrm{m\,s^{-1}}$.}

\subsection{Existence condition for resonance curves of gravity--capillary waves}
We aim to identify the conditions under which the resonance mechanism exists for a given wavenumber $k_y$. 
The analyses in \S~\ref{sec3} demonstrate that the resonance condition always holds for any value of $k_y$ in the case of gravity waves. However, this is not the case for gravity--capillary waves.

We assume that the resonance condition is satisfied at wavenumber $\bm k = (k_x, k_y)$:
\begin{align}
    \bm k \cdot \bm U=\Lambda(\bm k).
\end{align}
This expression is equivalent to
\begin{align}
    k_x U=\sqrt{g\sqrt{k_x^2+k_y^2}+\frac{\sigma}{\rho_w}\left(\sqrt{k_x^2+k_y^2}\right)^3}.\label{eq:gcwave_resonance}
\end{align}
We define a dimensionless Froude number $\gamma_0$, which is based on the pressure convection velocity $U$, critical wavenumber $k_0$ and gravitational acceleration $g$, as follows:
\begin{align}
    \gamma_0=\frac{U^2 k_0}{g} {=\frac{2U^2}{c_{\mathrm{min}}^2}}.
\end{align}
By setting $\kappa_x=k_x/k_0$ and $\kappa_y=k_y/k_0$, we can simplify~\eqref{eq:gcwave_resonance} as
\begin{align}
    \gamma_0  \kappa_x^2=\sqrt{\kappa_x^2+\kappa_y^2}+\left(\sqrt{\kappa_x^2+\kappa_y^2}\right)^3.\label{eq:gammakx}
\end{align}
Therefore, $\kappa_x^2$ is a positive root of the following cubic equation for $\xi$:
\begin{align}
    \xi^3+(3\kappa_y^2+2-\gamma_0^2)\xi^2+(1+\kappa_y^2)(1+3\kappa_y^2)\xi+\kappa_y^2(1+\kappa_y^2)^2=0.\label{eq:cubic_eqn}
\end{align}
The coefficients in \eqref{eq:cubic_eqn} are all real numbers, and we can obtain the properties of the roots using the discriminant. Let $\xi_i ( i=1,2,3)$ be the three roots of the cubic equation~\eqref{eq:cubic_eqn}. 
We note that \eqref{eq:cubic_eqn} has either three real roots or one real root and two complex conjugate roots.

We first consider the case when $\kappa_y\neq0$.
According to Vieta's formulas, we have
\begin{align}
    \xi_1\xi_2\xi_3=-\kappa_y^2(1+\kappa_y^2)<0.
\end{align}
Therefore, when $\xi_1$ and $\xi_2$ are complex conjugate roots, the only real root $\xi_3$ is negative, i.e., $\xi_3<0$. 
When all the roots are real numbers, $\xi_1\geq\xi_2>0$ and $\xi_3<0$. 
To determine the properties of the solutions, we next calculate the discriminant $\Delta_1$ for the cubic equation~\eqref{eq:cubic_eqn}:
\begin{align}
    \Delta_1
    =& (1 + \kappa_y^2)^2\gamma_0^2
    \left[4  \kappa_y^2 \gamma_0^4+ \left(1 - 18 \kappa_y^2 - 27 \kappa_y^4\right)\gamma_0^2  - 4\right].\label{eq:discriminant}
\end{align}
The cubic equation \eqref{eq:cubic_eqn} for $\xi$ has three real roots if and only if the discriminant satisfies $\Delta_1\geq0$.  Equation~\eqref{eq:cubic_eqn} has one negative root and two complex conjugate roots when the discriminant satisfies $\Delta_1<0$.

Based on \eqref{eq:discriminant}, we can infer that the sign of $\Delta_1$ is the same as that of $F_{\gamma}$:
\begin{align}
    F_{\gamma}(\gamma_0^2)=4 \kappa_y^2 \gamma_0^4+ \left(1 - 18  \kappa_y^2 - 27 \kappa_y^4\right)\gamma_0^2  - 4,
\end{align}
which can be interpreted as a quadratic function of $\gamma_0^2$.  The discriminant of the quadratic function $F_{\gamma}$ is always positive; hence, there exist two real roots $\psi_1$ and $\psi_2$ such that $F_{\gamma}(\psi_1)=F_{\gamma}(\psi_2)=0$.  
For simplicity, we assume that $\psi_1<\psi_2$.  Owing to Vieta's formula, we have $\psi_1\psi_2=-4<0$, and we obtain $\psi_1<0<\psi_2$. 
Therefore, we conclude that $F_\gamma(\psi)<0$ for any $\psi\in (0,\psi_2)$ and that $F_\gamma(\psi)>0$ for any $\psi\in (\psi_2,\infty)$.  By direct calculations, we obtain that 
\begin{align}
    \psi_2=\frac{-1+18\kappa_y^2+27\kappa_y^4+\sqrt{(1+\kappa_y^2)(1+9\kappa_y^2)^3}}{8\kappa_y^2}.
\end{align}
Hence, we have precise control over the sign of $\Delta_1$:
\begin{align}
    \Delta_1<0,\quad & \mathrm{when}\, \gamma_0<\sqrt{\psi_2},\\
    \Delta_1>0,\quad & \mathrm{when}\, \gamma_0>\sqrt{\psi_2}.
\end{align}
If the condition $\gamma_0<\sqrt{\psi_2}$ is satisfied, the cubic equation \eqref{eq:cubic_eqn} for $\xi$ has only one negative real root $\xi_1$. 
However, for $\xi_1$ to satisfy \eqref{eq:gammakx}, it must also be nonnegative, which is contradictory. 
As a result, for any $k_y$, the resonance condition \eqref{eq:gcwave_resonance} has no solution when $\gamma_0<\sqrt{\psi_2}$.
When $\gamma_0>\sqrt{\psi_2}$, we have the discriminant $\Delta_1>0$, and there exist two positive roots (i.e., $\xi_1$ and $\xi_2$) for the cubic equation~\eqref{eq:cubic_eqn}.  
Therefore, the resonance condition~\eqref{eq:gcwave_resonance} holds when the wavenumbers $\kappa_x$ are $\kappa_x=\sqrt{\xi_1}$ and $\kappa_x=\sqrt{\xi_2}$.
The function $\psi_2$ monotonically increases with respect to $\kappa_y$ when $\kappa_y$ is nonnegative. 
Hence, when $\kappa_y$ equals zero, $\psi_2$ obtains its minimum value of $\psi_2=4$.  In conclusion, the resonance curve does not exist in the wavenumber space if the Froude number $\gamma_0<\min(\sqrt{ \psi_2})=2$, i.e. $U^2/\sqrt{g\sigma/\rho_w}<2$.

\subsection{Characterisation of the resonance curve}\label{sec:4.2}
\begin{figure}
	\centering
         \captionsetup{width=1\linewidth, justification=justified}
        \includegraphics[width=0.7\textwidth]{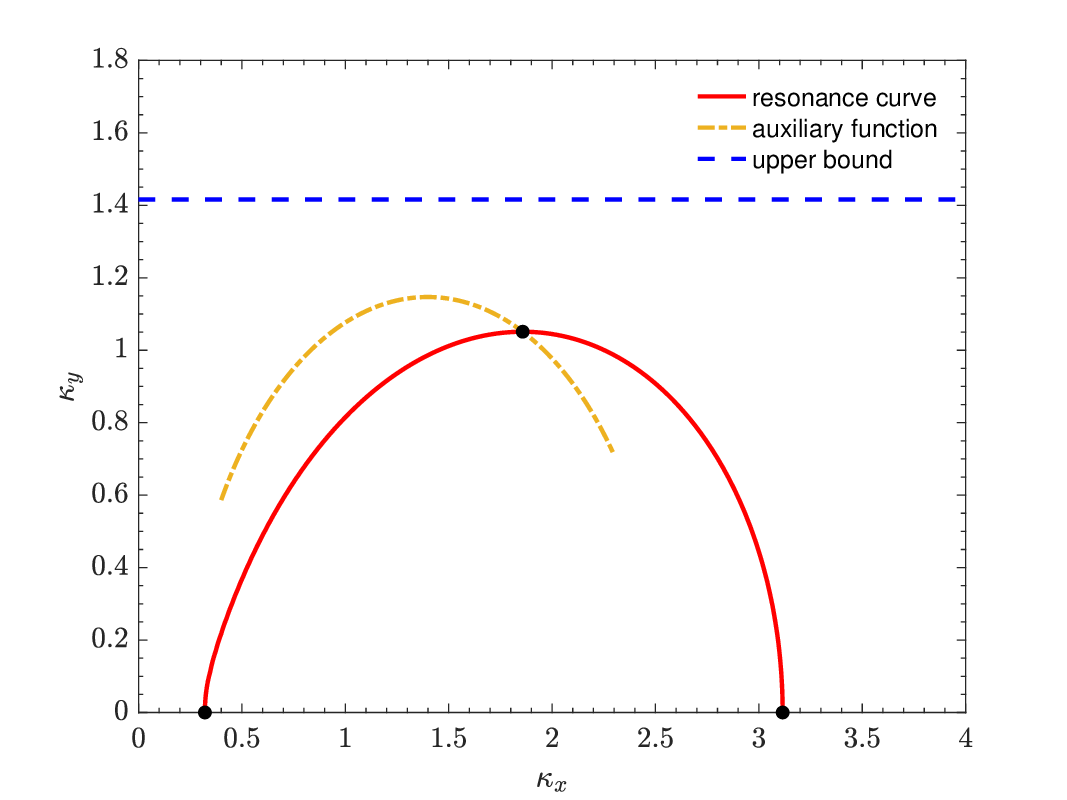}
	\caption{Illustration of the resonance curve.  The red solid line represents the resonance curve $\chi^*(\kappa_x,\kappa_y)=0$ based on \eqref{eq:gcwave_resonance}. The orange dashed line represents the auxiliary function defined in \eqref{eq:adj_eqn}. The dashed blue line denotes the theoretical upper bound, as defined in \eqref{eq:upperbound}, for $\gamma_0=3.44$.
    } \label{fig:resonance_curve}
\end{figure}

As illustrated in figure~\ref{fig:resonance_curve}, for gravity--capillary waves, the dimensionless resonance curve $\chi^*=0$ in the $(\kappa_x,\kappa_y)$ space can be characterised by three representative points: the two intersections between the resonance curve and the $\kappa_x$ axis and the maximum point $(\kappa_{x,M},\kappa_{y,M})$, where $\kappa_{y,M}$ is the largest value.
To further illustrate the shape of the resonance curve, we estimate the value of $\kappa_{y,M}$:
\begin{align}
    \kappa_{y,M}=\max(\kappa_y:\chi^*(\kappa_x,\kappa_y)=0).\label{eq:4.2:def_kappa_y_M}
\end{align}
We aim to determine an upper bound for  {$\kappa_y$} that satisfies the resonance condition.  Rearranging equation~\eqref{eq:gammakx}, we obtain
\begin{align}
    &(1+\kappa_y^2)(1+3\kappa_y^2)\kappa_x^2+\kappa_y^2(1+\kappa_y^2)^2=\kappa_x^4(\gamma_0^2-3\kappa_y^2-2-\kappa_x^2).\label{eq:4.2:deriv2}
\end{align}
The right-hand side of \eqref{eq:4.2:deriv2} has an upper bound due to the inequality of the arithmetic and geometric means.  Thus, we have
\begin{align}
     {\kappa_y^6\leq(1+\kappa_y^2)(1+3\kappa_y^2)\kappa_x^2+\kappa_y^2(1+\kappa_y^2)^2
    \leq\frac{4}{27}\left({\gamma_0^2-3\kappa_y^2-2}\right)^3.}
\end{align}
Therefore, for any $\kappa_y$ on the resonance curve $\chi^*(\kappa_x,\kappa_y)=0$, we obtain its upper bound as follows:
\begin{align}
     {\kappa_y\leq \frac{\sqrt[3]{2}}{\sqrt{3+3\sqrt[3]{4}}}\sqrt{\gamma_0^2-2}.}\label{eq:upperbound}
\end{align}
Here, we use the positivity of $(1+\kappa_y^2)(1+3\kappa_y^2)\kappa_x^2$  to estimate $\kappa_y$ without explicit knowledge of $\kappa_x$.

\begin{figure}
	\centering
        \captionsetup{width=1\linewidth, justification=justified}
        \includegraphics[width=0.7\textwidth]{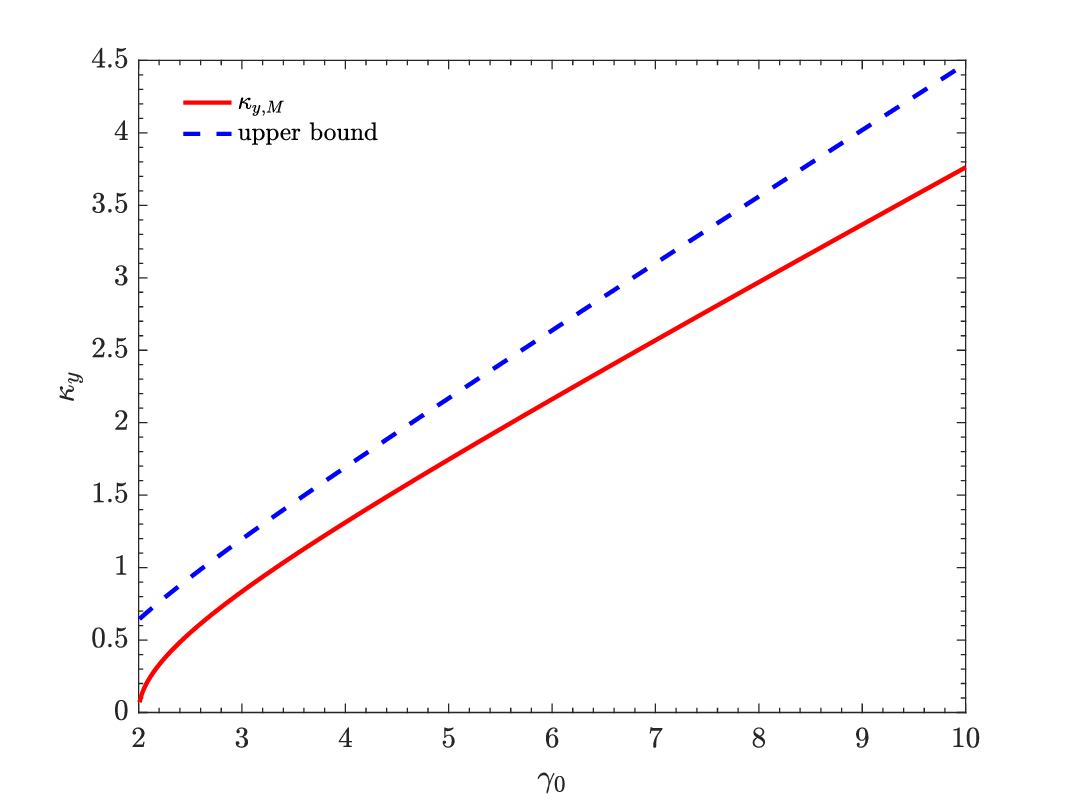}
	\caption{Variation in the maximum value $\kappa_{y,M}$ as a function of $\gamma_0$. The solid red line represents $\kappa_{y,M}$ solved according to~\eqref{eq:gammakx} and~\eqref{eq:adj_eqn}, and the dashed blue line denotes its theoretical upper bound, as defined in \eqref{eq:upperbound}.}\label{kym_gamma}
\end{figure}

We next determine the maximum value of $\kappa_y$ on the resonance curve $\chi^*(\kappa_x,\kappa_y)=0$.
We observe that the derivative of $\kappa_y$ with respect to $\kappa_x$ reaches zero at $\kappa_y=\kappa_{y,M}$.  
Taking the derivative of  {\eqref{eq:gammakx}} with respect to $\kappa_x$ and evaluating it at $\kappa_y=\kappa_{y,M}$, we obtain the following auxiliary function:
\begin{align}
    3\kappa_x^4 + 2(-\gamma_0^2 + 2 + 3\kappa_y^2)\kappa_x^2 + (1 + \kappa_y^2)(1 + 3\kappa_y^2) = 0,\label{eq:adj_eqn}
\end{align}
which is plotted in figure~\ref{fig:resonance_curve}  {for $\gamma_0=3.44$ when the convection velocity is $U=0.3\,\mathrm{m\,s^{-1}}$}.
We can obtain the value of $\kappa_{y,M}$ by solving  {\eqref{eq:gammakx}} and \eqref{eq:adj_eqn} simultaneously, which is the interception of the two curves plotted in figure~\ref{fig:resonance_curve}.   
Figure~\ref{kym_gamma} shows the variation in $\kappa_{y,M}$ and its upper bound, as defined in \eqref{eq:upperbound}, for different $\gamma_0$ values.  The results show that our estimation of the upper bound reasonably captures the variation in $\kappa_{y,M}$.

The points at which the resonance curve intersects the $\kappa_x$-axis can be directly calculated. 
At $\kappa_y=0$, we have the following solutions to the resonance curve $\chi^*(\kappa_x,\kappa_y)=0$: 
\begin{align}
    &\kappa_{x,L} = \frac{1}{2}\gamma_0-\frac{1}{2}\sqrt{\gamma_0^2-4},\label{eq:kappa_xl}\\
    &\kappa_{x,R} = \frac{1}{2}\gamma_0+\frac{1}{2}\sqrt{\gamma_0^2-4}.\label{eq:kappa_xr}
\end{align}
The resonance curve exists when $\gamma_0>2$ and the distance between the two intersections, $|\kappa_{x,L}-\kappa_{x,R}|=\sqrt{\gamma_0^2-4}$, increases as $\gamma_0$ increases.  
The area bounded by the resonance curve and the $x$-axis scales as $\gamma_0^2$, which also increases with increasing $\gamma_0$.

In summary, we present a quantitative estimation for the shape of the resonance curve for gravity--capillary waves.  
To accurately compute this curve, one must solve an implicit cubic equation. 
Our approach simplifies the expression for estimating the resonance curve height.
It also obtains the intersections between the resonance curve and the $\kappa_x$-axis.  
The results demonstrate that as the wind speed decreases (i.e., $\gamma_0$ decreases), the size of the domain bounded by the resonance curve and the $\kappa_x$-axis decreases.
When $\gamma_0$ is less than $2$, the variables $\kappa_{x,L}$ and $\kappa_{x,R}$, defined in~\eqref{eq:kappa_xl} and~\eqref{eq:kappa_xr}, are no longer real-valued. 
This finding implies that when the wind speed falls below a certain threshold, the resonance curve does not exist.  
In \S~\ref{sec:4.3} below, we investigate the wave evolution under such weak wind conditions when the wind speed is too low to trigger the initial stage of wind-wave generation.
The strong wind conditions are investigated in~\S~\ref{sec:4.4}.

\subsection{Weak wind conditions}\label{sec:4.3}
Following the previous discussion in \S~\ref{sec:4.2}, the resonance mechanism does not occur for gravity--capillary waves under weak wind conditions.
In such scenarios, the convection velocity of the air pressure fluctuations at the water surface is consistently lower than the phase velocity of the water waves for all wavenumbers.  
Therefore, the inequality $B<\Lambda$ in~\eqref{eq:F1complex} holds for any wavenumber $(k_x,k_y)$, and there is no singularity in the denominator of $F(k_x,k_y,t)$, as shown in~\eqref{eq:def_F}. Hence, $F(k_x,k_y,t)$ is bounded by a function that is independent of time $t$:
    \begin{align}
		F(k_x,k_y,t)&=\frac{\Lambda}{(B^2-\Lambda^2)^2}
		\left\{\frac{1}{2}B^2+\frac{3}{2}\Lambda^2-\Lambda(\Lambda+B)\cos[(B-\Lambda)t]\right.\notag\\
		&\left.-\Lambda(\Lambda-B)\cos[(\Lambda+B)t]-\frac{1}{2}(B^2-\Lambda^2)\cos(2\Lambda t)\right\}\notag\\
            &\leq\frac{4\Lambda^3}{(\Lambda^2-B^2)^2}.\label{eq:F_weak_wind_est}
    \end{align}
Therefore, function $G(k_y,t)$, i.e., the integral of $F(k_x,k_y,t)$ with respect to $k_x$, satisfies
\begin{align}
    G(k_y,t)\leq \int_0^\infty \frac{4\Lambda^3}{(\Lambda^2-B^2)^2} \mathrm{d}k_x.\label{eq:weak_est_G1}
\end{align}
According to the above equation, $G(k_y,t)$ does not increase over time but instead reaches a finite value.

\begin{figure}
    \centering
    \captionsetup{width=1\linewidth, justification=justified}
    \includegraphics[width=0.8\textwidth]{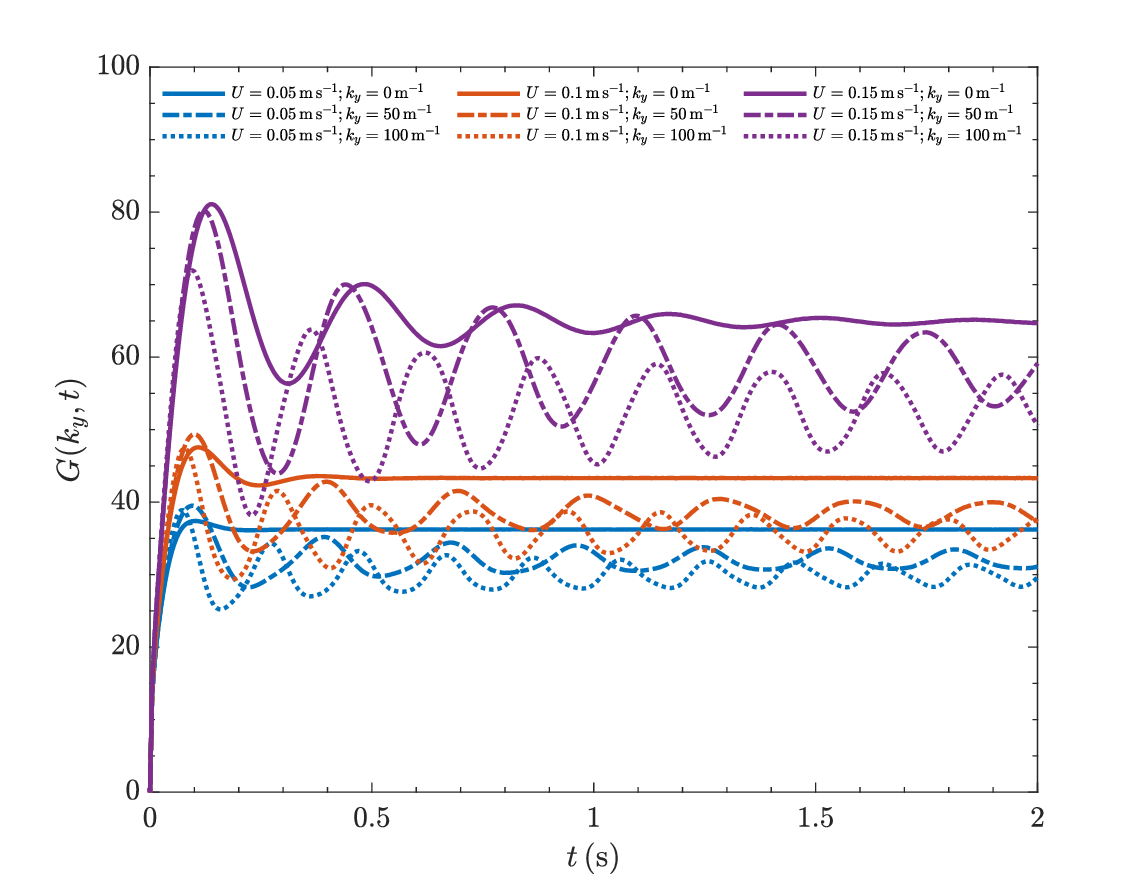}
	\caption{Time evolution of $G(k_y,t)$ for gravity--capillary waves under weak wind conditions for various parameter values. In all the cases, we set $g=9.8\,\mathrm{m\,s^{-2}}$, $\rho_w=10^3\,\mathrm{kg\,m^{-3}}$ and $\sigma=7\times10^{-2}\,\mathrm{kg\,s^{-2}}$.  The blue, red and purple lines correspond to the results obtained for $U=0.05\,\mathrm{m\,s^{-1}}$, $0.1\,\mathrm{m\,s^{-1}}$ and $0.15\,\mathrm{m\,s^{-1}}$, respectively. The solid lines, dash-dotted lines and dotted lines represent results for $k_y=0\,\mathrm{m^{-1}}$, $50\,\mathrm{m^{-1}}$ and $100\,\mathrm{m^{-1}}$, respectively.}\label{fig:weak_G_varying}
\end{figure}

Figure \ref{fig:weak_G_varying} shows examples of the temporal evolution of $G(k_y,t)$ for various values of $U$ and $k_y$.  We set $g=9.8\,\mathrm{m\,s}^{-1}$, $\rho_w=10^{3}\,\mathrm{kg\,m^{-3}}$ and $\sigma=7\times10^{-2}\,\mathrm{kg\,s^{-2}}$.  
The function $G(k_y,t)$ increases from zero over time and reaches a certain magnitude before beginning to oscillate.  
The amplitude of the oscillation in $G(k_y,t)$ decreases over time. 
At a fixed convection velocity of the pressure fluctuations $U$, the value of the function $G(k_y,t)$ decreases for larger values of $k_y$.
Given a fixed wavenumber $k_y$, the magnitude of $G(k_y,t)$ increases as the convection velocity $U$ increases.  
Thus, we can reasonably anticipate that higher wind speeds would induce larger surface elevation deformations. This expectation holds even when the wind speed is not sufficient to trigger wind-wave generation (the latter being a process in which the wave energy increases continuously over time).

Next, we aim to find an analytical expression that provides an upper bound for $G(k_y,t)$ for all values of time $t$, which will give us an estimation of the wave energy when the wind speed is too low to trigger the initial stage of wind-wave generation.

Under weak wind conditions, $\Lambda>B$ for any wavenumber $(k_x,k_y)$. We perform inequality scaling for the denominator of \eqref{eq:F_weak_wind_est} and obtain an upper bound estimation of the integral of $F(k_x,k_y,t)$ with respect to $k_x$. Our objective is to find a coefficient, denoted as $m\in(0,1)$, that does not rely on $k_x$ and satisfies the condition: 
\begin{align}
    m\Lambda^2 \geq B^2,\label{eq:sec4.4 est1}
\end{align}
for all values of $k_x$ and $k_y$. Therefore, we have $\Lambda^2-B^2>(1-m)\Lambda^2$. The right-hand side of \eqref{eq:F_weak_wind_est} is smaller than a power law function of $\Lambda$, whose integral can be estimated by a simple analytical solution.
 {Based on the inequality of arithmetic and geometric means, we have the following estimation:
\begin{align}
    \Lambda^2(k_x,k_y)=g|\bm k|+\sigma\rho_w^{-1}|\bm k|^3\geq2\sqrt{g\sigma\rho_w^{-1}}(k_x^2+k_y^2).\label{eq:sec4.4 est2}
\end{align}}
Because $B=Uk_x$, we combine \eqref{eq:sec4.4 est1} and \eqref{eq:sec4.4 est2} and obtain that \eqref{eq:sec4.4 est1} holds as long as $m$ satisfies
\begin{align}
    m\geq \frac{U^2}{2\sqrt{g\sigma\rho_w^{-1}}}= {\frac{U^2}{c_{\mathrm{min}}^2}}.\label{eq:sec4.2:range_m}
\end{align}
 {The existence of $m$ is ensured by the weak wind condition, i.e., $U<c_{\mathrm{min}}$. Because we have $c_{\mathrm{min}}=\sqrt{2\sqrt{g\sigma\rho_w^{-1}}}$, the inequality $U<c_{\mathrm{min}}$ is equivalent to $\frac{U^2}{2\sqrt{g\sigma\rho_w^{-1}}}<1$. Therefore, we can find an $m$ within the interval $(0,1)$ such that $m$ satisfies \eqref{eq:sec4.2:range_m}.}
Consequently, we can rescale the denominator of \eqref{eq:weak_est_G1} to eliminate the function $B$ and obtain the following result: 
\begin{align}
    G(k_y,t)
    &\leq \frac{4}{(1-m)^2}\int_0^\infty \frac{1}{\Lambda}\mathrm{d}k_x.
\end{align}
However, integrating $\Lambda^{-1}$ with respect to $k_x$ still does not lead to an analytical solution.  
To obtain an analytical upper bound estimation, we partition the integral domain $(0,\infty)$ into two parts: $(0,k_*)$ and $(k_*,\infty)$, where $k_*$ is an arbitrary positive number.  
To ensure that the integral remains well defined, we include only the gravity-related term in the partition $(0,k_*)$ and only the surface tension-related term in the partition $(k_*,\infty)$. 
This choice preserves sufficient physical information when determining the upper bound estimation.  
Hence, we obtain the following estimation for function $G(k_y,t)$:
\begin{align}
    G(k_y,t)&\leq \frac{4}{(1-m)^2}\left[\int_0^{k_*} \frac{1}{\sqrt{g\sqrt{k_x^2+k_y^2}}}\mathrm{d}k_x+ \int_{k_*}^\infty \frac{1}{\sqrt{\sigma\rho_w^{-1}\left(\sqrt{k_x^2+k_y^2}\right)^3}}\mathrm{d}k_x\right]\notag\\
    &\leq \frac{4}{(1-m)^2}\left[\int_0^{k_*} \frac{2^{\frac{1}{4}}}{\sqrt{g}(k_x+k_y)^\frac{1}{2}}\mathrm{d}k_x
    + \int_{k_*}^\infty \frac{2^{\frac{3}{4}}}{\sqrt{\sigma\rho_w^{-1}}\left(k_x+k_y\right)^\frac{3}{2}}\mathrm{d}k_x\right]\notag\\
    &\leq \frac{2^{\frac{13}{4}}}{(1-m)^2}\left[\frac{\sqrt{k_y+k_*}-\sqrt{k_y}}{\sqrt{g}}+\frac{\sqrt{2}}{\sqrt{\sigma\rho_w^{-1}}\sqrt{k_y+k_*}}\right].\label{eq:weak_wind_est_F_int1}
\end{align}
We use the inequality $k_x^2+k_y^2\geq(k_x+k_y)^2/2$ to obtain \eqref{eq:weak_wind_est_F_int1}, which rescales the integral of both the gravity- and surface tension-related terms to analytic solutions.  
Note that \eqref{eq:weak_wind_est_F_int1} holds for any value of $k_*$.  
We choose a specific value of $k_*$ that ensures that the right-hand side of \eqref{eq:weak_wind_est_F_int1} reaches a minimum when $k_y$ approaches zero.  
By calculations, we obtain that $k_*=\sqrt{2}k_0$.

Using the results of our previous analysis on the admissible range of $m$ (see \eqref{eq:sec4.2:range_m}), we choose $m=U^2/(2\sqrt{g\sigma\rho_w^{-1}})=\gamma_0/2$, and obtain:
\begin{align}
    G(k_y,t)\leq\frac{2^{\frac{21}{4}}}{\left(2-\gamma_0\right)^2}\left[\frac{\sqrt{k_y+\sqrt{2}k_0}-\sqrt{k_y}}{\sqrt{g}}+\frac{\sqrt{2}}{\sqrt{\sigma\rho_w^{-1}}\sqrt{k_y+\sqrt{2}k_0}}\right].\label{eq:weak_est_G2}
\end{align}
This estimation has useful physical implications. 
 {Equation~\eqref{eq:weak_est_G2} shows that the wave energy is bounded by a function that does not depend on time $t$ when the convection velocity of the air pressure fluctuations is below the  critical value, which is the minimum phase velocity of water waves}. 
Additionally, the saturated wave energy increases as the convection velocity increases. 
 {This relation between $G$ and $k_y$ can be supported by the upper bound estimation of $G$, derived in \eqref{eq:weak_est_G2}.  Both terms in parentheses on the right-hand side of \eqref{eq:weak_est_G2} decrease as $k_y$ increases.  Therefore, the upper bound of $G$ decreases as $k_y$ increases.}
The original analysis by~\citet{phillips1957generation} considered only the scenario in which the convection velocity is zero. 
Our approach here extends wave energy analysis to weak wind conditions (and the strong wind conditions, which are analysed in~\S~\ref{sec:4.4}), enabling a quantitative estimation of the dependency of the saturated wave energy on the convection velocity of the air pressure fluctuations at the water surface. 
\citet{perrard2019turbulent} investigated quasisteady surface elevation motions influenced by turbulent airflow based on the balance between the energy input from the airflow and the viscous dissipation in the water.
An experimental study by Paquier, Moisy \& Rabaud~(\citeyear{paquier2015surface}) reported that surface deformations can be saturated at low wind speeds.
Here, we present a physical mechanism for the generation of quasisteady surface elevation motions, even without the effect of viscous dissipation, when the airflow speed is inadequate to induce the resonance mechanism.

\begin{figure}
    \centering
    \captionsetup{width=1\linewidth, justification=justified}
    \includegraphics[width=0.99\textwidth]{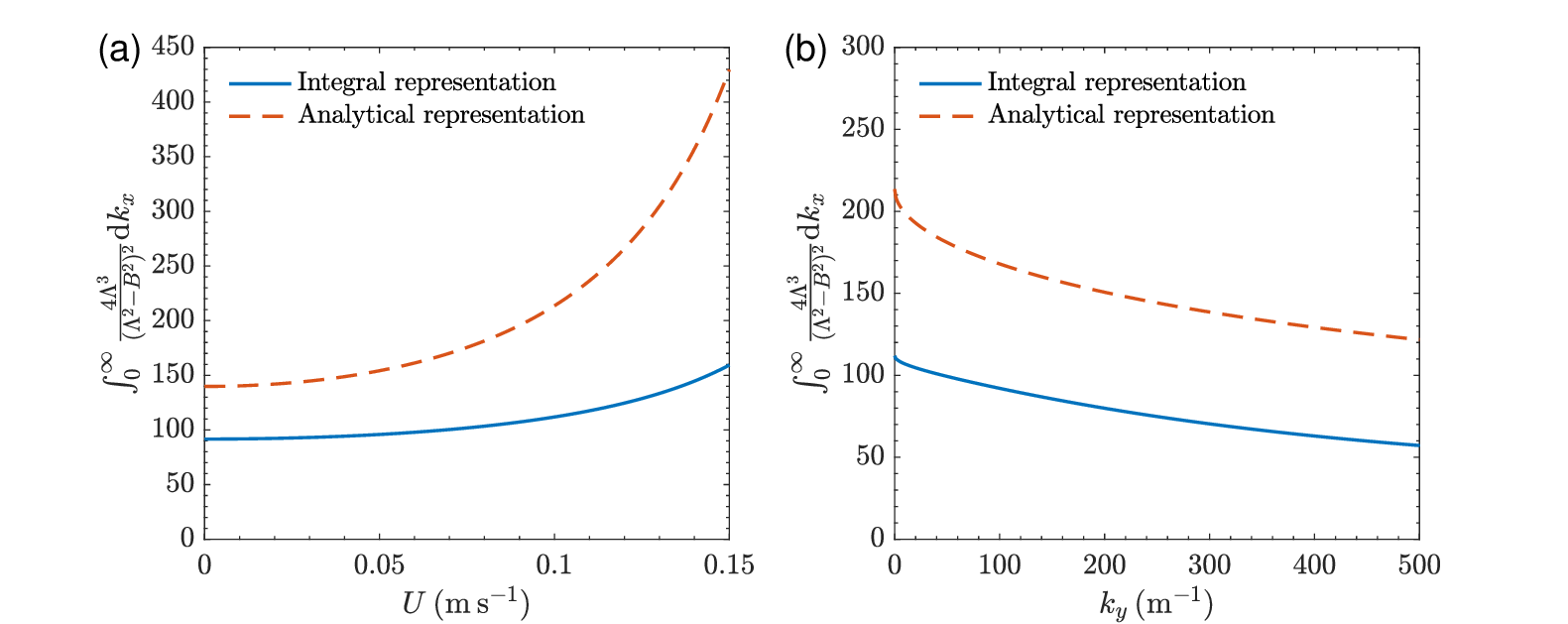}
    \caption{Evolution of $G(k_y,t)$ for gravity--capillary waves under weak wind conditions for different parameters. The blue solid lines denote the integral representation, i.e., the RHS of \eqref{eq:weak_est_G1}, and orange dashed lines show the analytical representation of the upper bound, i.e., the RHS of \eqref{eq:weak_est_G2}.
  Panel $(a)$ shows a plot of the representation functions versus the variable $U$, with $k_y=0$. Panel $(b)$ shows a plot of the representation functions versus the variable $k_y$, with $U=0.1\,\mathrm{m\,s^{-1}}$. In all cases, we choose $g=9.8\,\mathrm{m\,s^{-2}}$, $\rho_w=10^3\,\mathrm{kg\,m^{-3}}$ and $\sigma=7\times10^{-2}\,\mathrm{kg\,s^{-2}}$.}
 \label{fig:weak_G_compare}
\end{figure}

In figure~\ref{fig:weak_G_compare}, we present numerical results to support the above estimate of $G(k_y,t)$. The blue solid lines in the figure~\ref{fig:weak_G_compare} represent the integral representation of the right-hand side of \eqref{eq:weak_est_G1}, while the red dashed lines depict the analytical representation, that is, the right-hand side of \eqref{eq:weak_est_G2}. The results show that the estimation using the analytical representation defined in \eqref{eq:weak_est_G1} reasonably captures the characteristics of the upper bound of the function $G(k_y,t)$ under weak wind conditions.

 {Under weak wind conditions, the wave energy eventually reaches a saturation value because the resonance condition for linear growth does not exist.  Numerical examples confirm that the saturation energy increases with the wind speed, and our derived upper bound for $G$ under weak wind conditions approaches infinity when $U$ approaches $c_{\mathrm{min}}$.  This result indicates that, under weak wind conditions, as the wind speed approaches the minimum phase velocity of gravity--capillary water waves, $G$ can be extremely large. This result suggests a continuous transition from weak to strong wind conditions. Furthermore, it is also important to emphasise that our present analysis applies only to the initial stage of wind-wave generation. The governing equations for wave motion in this stage assume Taylor’s frozen turbulence hypothesis. Over longer time scales, time decorrelations of pressure fluctuations occur, and the wave evolution depends on the space--time correlation of air turbulence~\citep{li2022principal}.}

\subsection{Strong wind conditions}\label{sec:4.4}
In this subsection, we analyse wave evolution under strong wind conditions,  namely,  when the convection velocity of the air pressure fluctuations exceeds the minimum phase velocity of the water waves.
As stated in the previous discussions, such circumstances are met when the nondimensional Froude number satisfies $\gamma_0 > 2$; hence, resonance occurs for wavenumbers within the range of  {$k_y \leq k_{y,M}$}.
Here, $k_{y,M}$ represents the maximum wavenumber for which resonance occurs and is expressed as $k_{y,M} = k_0\kappa_{y,M}$, where $\kappa_{y,M}$ is defined in \eqref{eq:4.2:def_kappa_y_M}.  
We employ a technique similar to the method proposed in \S~\ref{sec3} to analyse the wave energy evolution for gravity--capillary waves next.
{We decompose the evolution of $G(k_y,t)$ into a linear growth term $M(k_y,t)$ and a residual term $N(k_y,t)$. Our numerical results show that $M(k_y,t)$ accounts for the leading-order growth of $G(k_y,t)$ in most cases.}

\subsubsection{Analysis in the complex plane}
Similar to the analysis of gravity waves, for gravity--capillary waves we define the analytical continuation of $F$ as the following complex-valued function $F_2(z,k_y,t)$:
	\begin{align}
		F_2(z,k_y,t)=&\frac{\Lambda}{(B^2-\Lambda^2)^2}
		\left\{\frac{1}{2}B^2+\frac{3}{2}\Lambda^2-\Lambda(\Lambda+B)\exp[\mathrm{i}(\Lambda-B)t]\right.\notag\\
		&\left.-\Lambda(\Lambda-B)\exp[\mathrm{i}(\Lambda+B)t]-\frac{1}{2}(B^2-\Lambda^2)\exp(2\mathrm{i}\Lambda t)\right\}.\label{eq:F1complex_gcwave}
	\end{align}
Here, the complex-valued functions $B$ and $\Lambda$ are defined as
\begin{align}
    B(z)&=Uz,\\
    \Lambda(z)&=\sqrt{g\sqrt{z^2+k_y^2}+\sigma\rho_w^{-1}\left(\sqrt{z^2+k_y^2}\right)^3}.
\end{align}
The term $\exp[\mathrm{i}(\Lambda-B)t]$ in \eqref{eq:F1complex_gcwave} ensures that $F_2(z,k_y,t)$ approaches zero for large values of $|z|$ in the first quadrant of the complex plane for gravity--capillary waves.

\begin{figure}
	\centering
\includegraphics[width=0.45\textwidth]{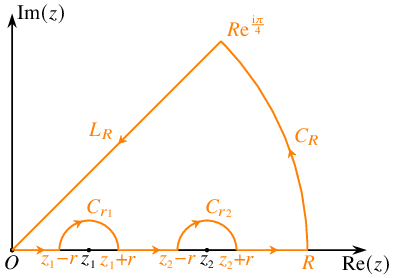}
	\caption{Illustration of the closed loop for integration, as outlined in ~\eqref{eq:loop2}.\label{fig:illus loop2}}
\end{figure}

When the spanwise  {wavenumber} $k_y$ is less than the maximum wavenumber for which the resonance occurs, i.e., $k_{y,M}$, two points $(k_{x1},k_y)$ and $(k_{x2},k_y)$ exist that satisfy the resonance condition.
We consider a closed loop $\Gamma$ in the complex plane, as depicted in figure~\ref{fig:illus loop2}, which is defined as $\Gamma=[0, z_1-r]+C_{r1}+[z_1+r, z_2-r]+C_{r2}+[z_2+r, R]+C_R+L_R$.
 {$z_1$ and $z_2$ are the two roots of $B(z)-\Lambda(z)=0$.}
Here, $C_{r1}( {\theta})=z_1+re^{\mathrm{i}(\pi-\theta)}$, with $0\leq \theta \leq \pi$, is a clockwise circular path around $z_1$ with a radius of $r$; 
$C_{r2}( {\theta})=z_2+re^{\mathrm{i}(\pi-\theta)}$, with $0\leq \theta \leq \pi$, is a clockwise circular path around $z_2$ with a radius of $r$;
$C_R(\theta)=Re^{\mathrm{i}\theta}$, with $0\leq \theta \leq \pi/4$, is an anticlockwise circular path around 0 with a radius of $R$; and $L_R(\theta)=Re^{\mathrm{i}\pi/4}(1-\theta)$, with $0\leq\theta\leq 1$, is a straight path pointing towards the origin.
Because the function $F_2$ has no singularities inside the closed loop $\Gamma$, the integral of the function $F_2(z,k_y,t)$ with respect to $z$ along the contour $\Gamma$ is equal to zero, which can be expressed as
\begin{align}
	\left(\int_0^{z_1-r}+\int_{C_{r1}}+\int_{z_1+r}^{z_2-r}+\int_{C_{r2}}
 +\int_{z_2+r}^R+\int_{C_R}+\int_{L_R}\right)F_2(z,k_y,t)\mathrm{d}z=0.\label{eq:loop2}
\end{align}
Taking the limits of $R\rightarrow\infty$ and $r\rightarrow 0$, the integration of $F_2$ along the real positive axis can be calculated as
\begin{align}
	\int_0^{\infty}F(k_x,k_y,t) \mathrm{d}k_x=&\lim_{R\rightarrow\infty}\int_{0}^{R}F(k_x,k_y,t) \mathrm{d}k_x\notag\\
	=&\lim_{R\rightarrow\infty}\lim_{r\rightarrow0}\mathrm{Re}\left[\left(\int_0^{z_1-r}+\int_{z_1+r}^{z_2-r}+\int_{z_2+r}^{R}\right)F_2(z,k_y,t)\mathrm{d}z\right]\notag\\
	=&\lim_{R\rightarrow\infty}\lim_{r\rightarrow0}\mathrm{Re}\left[\left(-\int_{C_{r1}}-\int_{C_{r2}}-\int_{C_R}-\int_{L_r}\right)F_2(z,k_y,t)\mathrm{d}z\right]\notag\\
	=&-\lim_{R\rightarrow\infty}\mathrm{Re}\left[\int_{C_R}F_2(z) \mathrm{d}z\right]
	-\lim_{r\rightarrow0}\mathrm{Re}\left[\int_{C_{r1}}F_2(z) \mathrm{d}z\right]\notag\\
	&-\lim_{r\rightarrow0}\mathrm{Re}\left[\int_{C_{r2}}F_2(z) \mathrm{d}z\right]-\lim_{R\rightarrow\infty}\mathrm{Re}\left[\int_{C_{L_R}}F_2(z) \mathrm{d}z\right].\label{eq:4.35}
\end{align}
Similar to the above analysis for gravity waves, the integration of $F(k_x,k_y,t)$ with respect to $k_x$ along the positive $x$-axis is equivalent to the integration of the complex function $F_2(z,k_y,t)$ with respect to the complex value $z$ along the paths $C_{r1}$, $C_{r2}$, ${C_R}$, and ${L_R}$.  
Because the magnitudes of $B$ and $\Lambda$ behave asymptotically as $|B|\sim |z|$ and $|\Lambda|\sim |z|^{3/2}$ when $|z|\rightarrow\infty$, the function $F_2(z,k_y,t)$, defined in \eqref{eq:F1complex_gcwave}, behaves as $|F_2(z,k_y,t)|\sim |z|^{-3/2}$ for large values of $|z|$ when $z$ is located in the first quadrant of the complex $z$-plane.  Therefore, the integration of $F_2$ along the path $C_R$ becomes zero as $R$ approaches infinity.
Hence, we next focus on the integration of $F_2(z,k_y,t)$ along the paths $C_{r1}$, $C_{r2}$, and $C_{L_R}$.

\subsubsection{Calculation of the integral of $F_2$ along the paths $C_{r1}$ and $C_{r2}$}
The integral of $F_2$ along the semicircular paths $C_{r1}$ and $C_{r2}$ can be calculated using the residue theorem as follows:
\begin{align}
	\lim_{r\rightarrow0}\int_{C_{r1}}F_2(z) \mathrm{d}z&=-\pi\mathrm{i}\mathrm{Res}(F_2,z_{1}),\notag\\
    \lim_{r\rightarrow0}\int_{C_{r2}}F_2(z) \mathrm{d}z&=-\pi\mathrm{i}\mathrm{Res}(F_2,z_{2}).
\end{align}
The varibales $z_1$ and $z_2$ are two real-valued solutions for $z$ to the following resonance condition equation:
\begin{align}
    Uz=\sqrt{g\sqrt{z^2+k_y^2}+\sigma\rho_w^{-1}\left(\sqrt{z^2+k_y^2}\right)^3}.
\end{align}
Both singularity points (i.e., $z=z_1$ and $z=z_2$) are first-order poles of the complex-valued function {$F_2(z,k_y,t)$}. The residues at $z=z_i$ for $i=1,2$ are defined as 
\begin{align}
    \mathrm{Res}(F_2,z_i)=\lim_{z\rightarrow z_i} F_2(z)(z-z_i).\label{eq:res:F2}
\end{align}
To compute these residues, we follow the same perturbation method introduced in~\S~\ref{sec:3.2.3}. We begin by calculating the Taylor expansions of $B-\Lambda$ around $z-z_1$ and $z-z_2$, respectively. These expansions are then substituted into \eqref{eq:res:F2}, yielding
\begin{align}
    \mathrm{Res}( {F_2},z_1) &=-\frac{ \mathrm{i}U^2 z_1\sqrt{z_1^2+k_y^2}t}
    {g+3\sigma\rho_w^{-1}(z_1^2+k_y^2)-2U^2\sqrt{z_1^2+k_y^2}},\notag\\
        \mathrm{Res}( {F_2},z_2) &=-\frac{ \mathrm{i}U^2 z_2\sqrt{z_2^2+k_y^2}t}
    {g+3\sigma\rho_w^{-1}(z_2^2+k_y^2)-2U^2\sqrt{z_2^2+k_y^2}}.
\end{align}
Hence, the integral of $F_2$ along two paths $C_{r1}$ and $C_{r2}$ is
\begin{align}
    \int_{C_{r1}+C_{r2}}F_2(z,k_y,t)dz=&-\frac{\pi U^2 z_1\sqrt{z_1^2+k_y^2}t}
    {g+3\sigma\rho_w^{-1}(z_1^2+k_y^2)-2U^2\sqrt{z_1^2+k_y^2}}\notag\\
    &-\frac{\pi U^2 z_2\sqrt{z_2^2+k_y^2}t}
    {g+3\sigma\rho_w^{-1}(z_2^2+k_y^2)-2U^2\sqrt{z_2^2+k_y^2}}.\label{eq:4.39}
\end{align}
Notably, this result indicates that the integral of $F_2$ along the paths $C_{r1}$ and $C_{r2}$ is a linear function of time $t$.
 {At $k_y=k_{y,M}$, the two roots $z_1$ and $z_2$ coincide and are denoted as $z_M$. In other words, the integral in~\eqref{eq:4.39} can then be evaluated with $z_1=z_2=z_M$.}
Next, we analyse the integral of $F_2$ along the path $L_R$.

\subsubsection{Calculation of the integral of $F_2$ along path $L_R$}

Next, we consider the integral of $F_2$ along path $L_R$, where the variable $z$ can be parameterised as a function of the real variable $r$, such that $z(r) = r\exp(\mathrm{i}\pi/4)$ with $0\leq r\leq R$.  Therefore, we have
\begin{align}
	\int_{L_R} F_2(z)\mathrm{d}z&=\exp\left(\frac{\mathrm{i}\pi}{4}\right)\int_R^0 F_2(z(r))\mathrm{d}r\notag\\
	&=-\frac{1+\mathrm{i}}{\sqrt{2}}\int_0^R\frac{K(r,t)}{J(r)}\mathrm{d}r.\label{eq:LR1}
\end{align}
In \eqref{eq:LR1}, the numerator $K(r,t)$ and denominator $J(r)$ are
\begin{align}
	K(r,t)=&\Lambda_0(r)\left\{\frac{1}{2}B_0(r)^2+\frac{3}{2}\Lambda_0(r)^2-\Lambda_0(r)[\Lambda_0(r)+B_0(r)]\exp[\mathrm{i}(\Lambda_0(r)-B_0(r))t]\right.\notag\\
	&-\Lambda_0(r)[\Lambda_0(r)-B_0(r)]\exp[\mathrm{i}(B_0(r)+\Lambda_0(r))t]\notag\\
	&\left.-\frac{1}{2}\left[B_0(r)^2-\Lambda_0(r)^2\right]\exp[2\mathrm{i}\Lambda_0(r) t]\right\}, \label{eq:expr_K}\\
	J(r)=&\left[B_0(r)^2-\Lambda_0(r)^2\right]^2,\label{eq:expr_J}
\end{align}
respectively, and $B_0(r)$ and $\Lambda_0(r)$ are defined as 
\begin{align}
    B_0(r)&=\frac{1+\mathrm{i}}{\sqrt{2}}Ur,\notag\\
    \Lambda_0(r)&=\sqrt{g\sqrt{\mathrm{i}r^2+k_y^2}+\sigma\rho_w^{-1}\left(\sqrt{\mathrm{i}r^2+k_y^2}\right)^3}.
\end{align}
 {We note that the definitions of $B_0(r)$ and $\Lambda_0(r)$ are based on the angle between $L_R$ and the $x$-axis being $\pi/4$.  However, this choice of angle does not influence the real part of the integral of $F_2$ along $L_R$ because the contribution from the integral along $C_R$ vanishes as $R\rightarrow\infty$.  This phenomenon has been verified through numerical integrations for different angles.}

\subsubsection{Properties of the function G}

Based on~\cref{eq:4.35,eq:4.39,eq:LR1} and take the limit $R\rightarrow\infty$, we can obtain the following expression for $G(k_y,t)$, defined in \eqref{eq:def G}:
\begin{align}
    G(k_y,t)=M(k_y,t)+N(k_y,t).
\end{align}
 {Here, the functions $M(k_y,t)$ and $N(k_y,t)$ are
\begin{align}
    M(k_y,t)=&\frac{\pi z_1\sqrt{z_1^2+k_y^2}U^2 t}
    {g+3\sigma\rho_w^{-1}\left(z_1^2+k_y^2\right)-2U^2\sqrt{z_1^2+k_y^2}}+\frac{\pi z_2\sqrt{z_2^2+k_y^2}U^2 t}
    {g+3\sigma\rho_w^{-1}\left(z_2^2+k_y^2\right)-2U^2\sqrt{z_2^2+k_y^2}},\label{eq:expr_M}\\
    N(k_y,t)=&\mathrm{Re}\left[\frac{1+\mathrm{i}}{\sqrt{2}}\int_0^\infty\frac{K(r,t)}{J(r)}\mathrm{d}r\right],\label{eq:expr_N}
\end{align}
where the expressions of the numerator $K(r,t)$ and the denominator $J(r)$ on the right-hand side of \eqref{eq:expr_N} are given in \eqref{eq:expr_K} and \eqref{eq:expr_J}, respectively. Due to its complexity, we do not explicitly write the complete expression of $N(k_y,t)$ here.
{The function $M(k_y,t)$ is the leading-order term of $G(k_y,t)$ in most cases, and $N(k_y,t)$ is the residual.}
The function $N(k_y,t)$ can be computed through numerical integration using relationships in \eqref{eq:expr_K}, \eqref{eq:expr_J} and \eqref{eq:expr_N}.}
 {Unlike the gravity wave case, in \eqref{eq:expr_M}, $M(k_y,t)$ does not exhibit an explicit scaling behaviour with the convection velocity $U$.  Assume that the two roots are ordered such that $z_1<z_2$.  The smaller root $z_1$ (corresponding to the smaller wavenumber) can be treated as the `gravity approximation' of the solution, while the larger root $z_2$ (larger wavenumber) represents the `capillary approximation' of the solution. {Preliminary numerical test indicates that the growth rate of $M(k_y,t)$ associated with $z_1$ (the first term on the right-hand side of \eqref{eq:expr_M}) is negative and scales as $U^n$ with $n\in[-2,-1]$.  The scaling behaviour is similar to the gravity wave case (see~\eqref{eq:q_ky_scaling}). On the other hand, the growth rate of $M(k_y,t)$ related to $z_2$ is postive and scales as $U^2$ when $|k_y|\ll z_2$, reflecting the capillary effect.}}
 
In the gravity wave case, we demonstrated that $N(k_y,t)$ is an oscillating function with a bounded amplitude. 
However, in the gravity--capillary wave case, due to the high-order nonlinearity term in the variable $\Lambda$, the residual term $N(k_y,t)$ both oscillates and increase linearly over time, but its magnitude is lower than that of the leading-order term $M(k_y,t)$. 
A special case in which $N(k_y,t)$ does not increase over time occurs when $g=0$ and $k_y=0$. 
Numerical examples are provided below to support the above theoretical findings.

\begin{figure}
	\centering
         \captionsetup{width=1\linewidth, justification=justified}
        \includegraphics[width=1\textwidth,trim={1cm 0 1.2cm 0},clip]{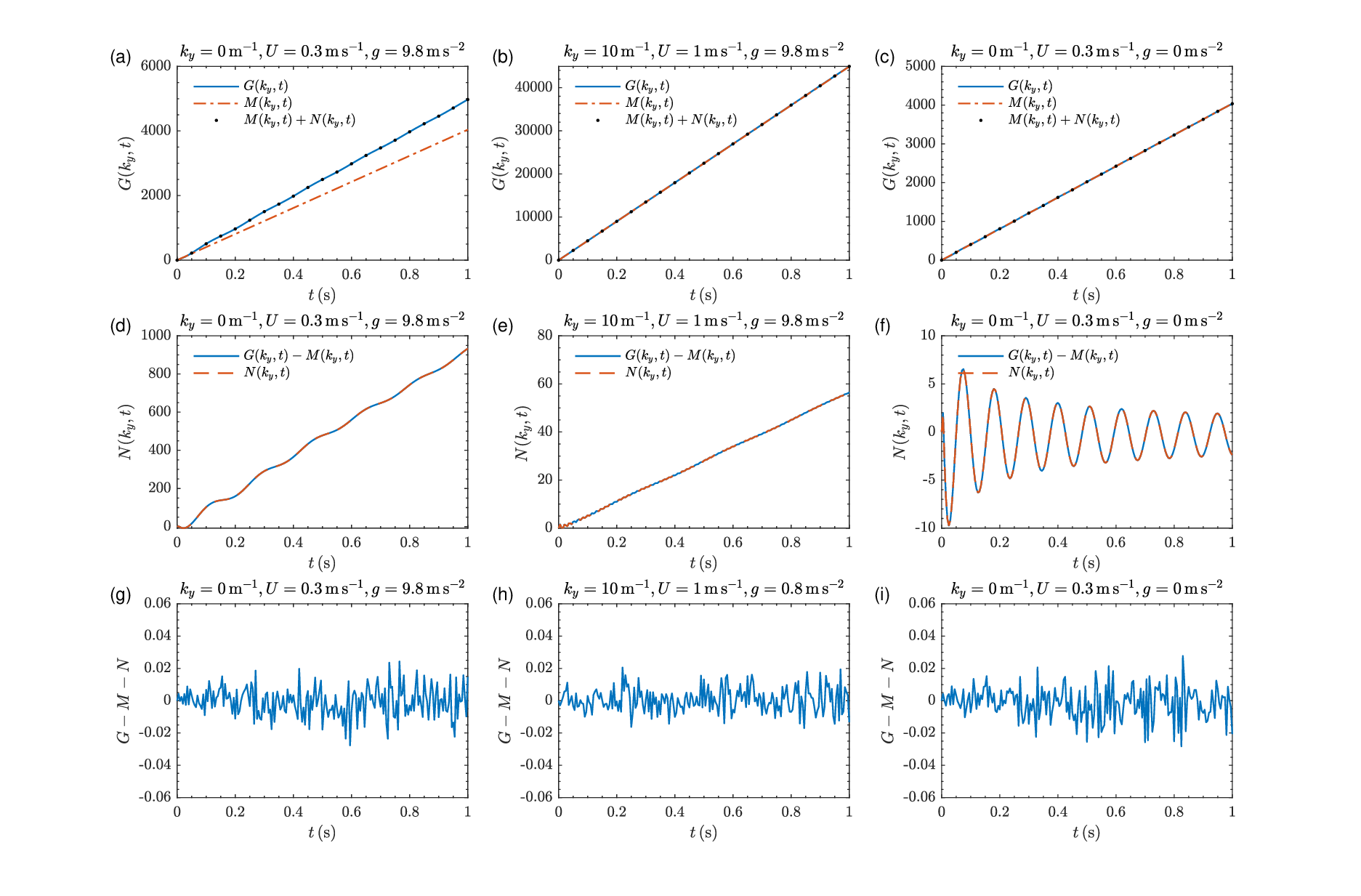}
	\caption{ {Numerical examples of the calculation of $G(k_y,t)$, with the results for $k_y=0\,\mathrm{m^{-1}}$, $U=0.3\,\mathrm{m\,s^{-1}}$ and $g=9.8\,\mathrm{m\,s^{-2}}$ shown in the left column (panels a, d, and g), the results for $k_y=10\,\mathrm{m^{-1}}$, $U=1\,\mathrm{m\,s^{-1}}$ and $g=9.8\,\mathrm{m\,s^{-2}}$ shown in the middle column (panels d, e, and h), and the results for $k_y=0\,\mathrm{m^{-1}}$ and $g=0\,\mathrm{m\,s^{-2}}$ shown in the right column (panels b, d, and f). The surface tension $\sigma$ and water density $\rho_w$ are set to $\sigma=7\times 10 ^{-2}\,\mathrm{kg\,s^{-2}}$ and $\rho_w=10^3\,\mathrm{kg\,m^{-3}}$, respectively. Panels (a), (b), and (c): $G(k_y,t)$, with the blue lines indicating the numerical integration results, the orange lines indicating the linear growth component $M(k_y,t)$, and the black dots indicating the sum of the linear growth component $M(k_y,t)$ and the oscillatory component $N(k_y,t)$. Panels (d), (e), and (f): the orange lines represent the oscillatory component $N(k_y,t)$, and the blue lines represent $G(k_y,t)-M(k_y,t)$. Panels (g), (h), and (i): numerical errors in the calculations, with the blue lines representing $G(k_y,t)-M(k_y,t)-N(k_y,t)$.}
    } \label{compare_gravity_capillary_waves}
\end{figure}

 {Figure \ref{compare_gravity_capillary_waves} shows numerical examples of calculating the function $G(k_y,t)$ with $M(k_y,t)$ and $N(k_y,t)$ for a gravity--capillary wave with different convection velocity $U$ (i.e., $U=0.3\,\mathrm{m\,s^{-1}}$ and $U=1\,\mathrm{m\,s^{-1}}$) and a capillary wave (a limiting case when $g=0\,\mathrm{m\,s^{-2}}$). 
The relationship among $G(k_y,t)$, $M(k_y,t)$ and $N(k_y,t)$ and their temporal evolution properties are visualised in panels (a), (b) and (c) of figure~\ref{compare_gravity_capillary_waves}.
The results confirm that the function $G(k_y, t)$ predominantly increases linearly over time and that the function $M(k_y, t)$ captures the leading-order behaviour of $G(k_y, t)$.  In panels (a) and (b), the convection velocity is $U=0.3\,\mathrm{m\,s^{-1}}$ and $U=1\,\mathrm{m\,s^{-1}}$, respectively, and in both cases, the strong wind condition is satisfied: $U>c_{\mathrm{min}}=0.23\,\mathrm{m\,s^{-1}}$. For the larger convection velocity case (i.e., panel b), the magnitude of $M(k_y,t)$ is closer to that of $G(k_y,t)$.
The time evolution of the function $N(k_y,t)$, which is the oscillatory component of $G(k_y,t)$, is shown in panels (c) and (d) of figure \ref{compare_gravity_capillary_waves}. 
In general, $N(k_y,t)$ also increases over time and oscillates, and the magnitude of $N(k_y,t)$ is smaller than that of $M(k_y,t)$, confirming that $M(k_y,t)$ represents the leading-order behaviour of the temporal evolution of $G(k_y,t)$. 
The special case of $g=0$ and $k_y=0$ is shown in the right column of figure~\ref{compare_gravity_capillary_waves}. 
In this case, the residual term $N(k_y,t)$ does not increase and instead only oscillates over time, and its amplitude decreases gradually.  
Panels (e) and (f) of figure \ref{compare_gravity_capillary_waves} show the numerical error for the integration of $G(k_y,t)$, calculated as $G(k_y,t)-M(k_y,t)-N(k_y,t)$. 
The relative error between $G(k_y,t)$ and $M(k_y,t)+N(k_y,t)$ normalised by $G(k_y,t)$ is less than $4\times10^{-5}$, which confirms the accuracy of the numerical integrations.}

\begin{figure}
	\centering
         \captionsetup{width=1\linewidth, justification=justified}
        \includegraphics[width=1\textwidth,trim={1cm 0 1.2cm 0},clip]{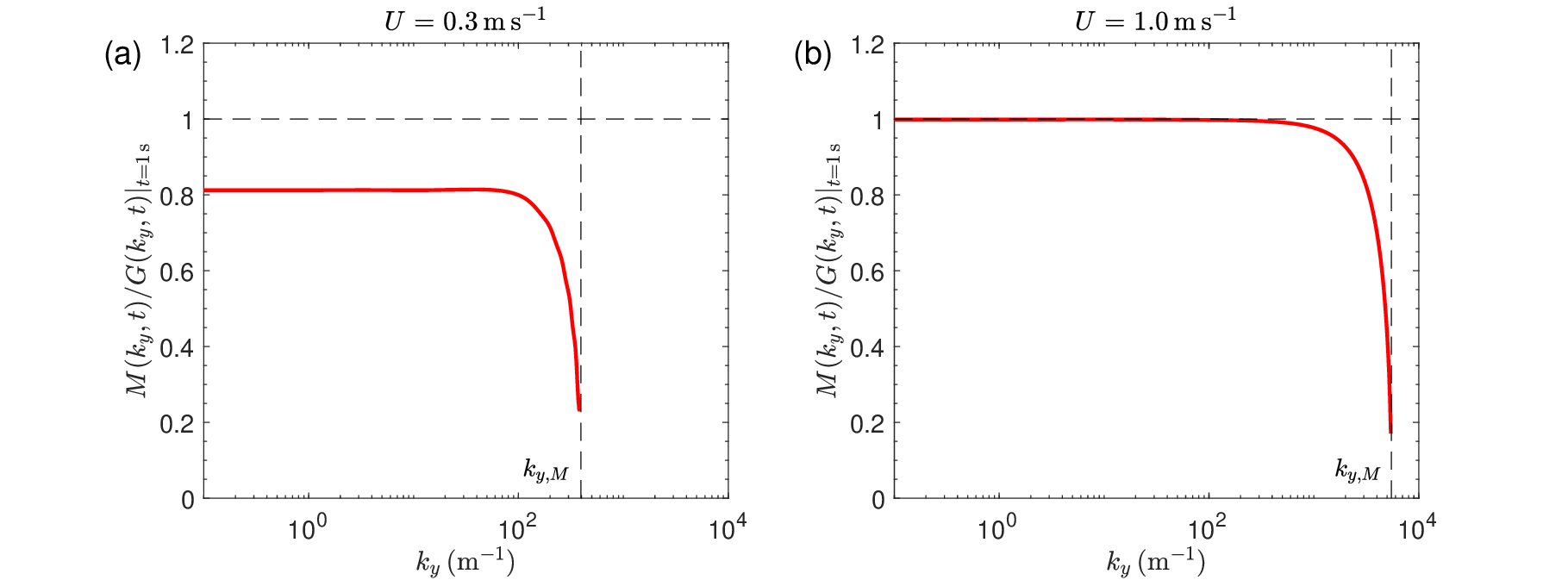}
	\caption{ {Two numerical examples of the ratio $M(k_y,t)/G(k_y,t)$ at $t=1\,\mathrm{s}$ with respect to wavenumber $k_y$.  {In each panel, the vertical dashed line indicates the location where $k_y=k_{y,M}$, and the horizonal dashed line represents the value 1. }}
    } \label{two_example_MG}
\end{figure}

\begin{figure}
	\centering
         \captionsetup{width=1\linewidth, justification=justified}
        \includegraphics[width=0.65\textwidth,]{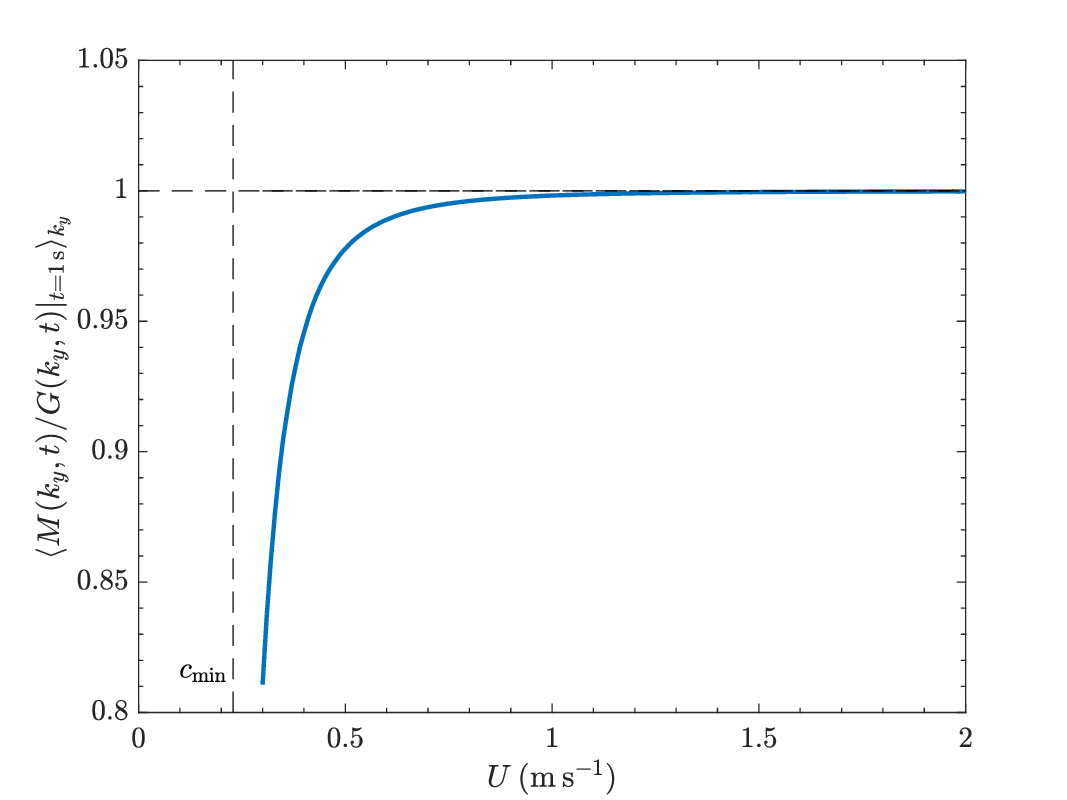}
	\caption{ {Ratio of $M(k_y,t)/G(k_y,t)$ at $t=1\,\mathrm{s}$, averaged over $k_y$, as a function of the convection velocity $U$. {The vertical dashed line indicates the location where $U=c_{\mathrm{min}}$, and the horizonal dashed line represents the value 1. }}
    } \label{ratio_MG_varying_U}
\end{figure}

 {To better understand the relative contributions of $M(k_y,t)$ and $N(k_y,t)$ to $G(k_y,t)$, we calculate the ratio of $M(k_y,t)/G(k_y,t)$ at $t=1\,\mathrm{s}$ for different $k_y$ and $U$. \Cref{two_example_MG} illustrates two numerical examples of $M(k_y,t)/G(k_y,t)|_{t=1\,\mathrm{s}}$ varying with the spanwise wavenumber $k_y$.  In both cases, the magnitude of $M(k_y,t)$ is more than one half of the magnitude of $G(k_y,t)$ for $k_y$ in the range of $(0,0.8k_{y,M})$.  For the higher wind velocity case (\cref{two_example_MG}\textit{b}), $M(k_y,t)/G(k_y,t)|_{t=1\,\mathrm{s}}$ is above $0.998$ for a wide range of $k_y$.  
\Cref{ratio_MG_varying_U} plots the ratio of $M(k_y,t)/G(k_y,t)$ at $t=1\,\mathrm{s}$, averaged over $k_y\in[0,100]\,\mathrm{m^{-1}}$, and illustrates that this value approaches one as the convection velocity $U$ increases.
{Therefore, $M(k_y,t)$ is a reasonable representation of the leading-order linear behaviour of $G(k_y,t)$ in most cases.}
}

\subsubsection{Overall increase in the surface elevation variance $\langle\eta^2\rangle$ over time}

When the spanwise wavenumber $k_y$ satisfies $k_y<k_{y,M}$, the resonance condition can be met at two points of $k_x$ for a fixed value of $k_y$. However, when $k_y$ is larger than the critical value $k_{y,M}$, no resonance occurs.  Hence, we partition the first quadrant of the wavenumber space into two parts: $\Omega_1=\{(k_x,k_y)\in[0,\infty)\times {[}0,k_{y,M})\}$ and $\Omega_2=\{(k_x,k_y)\in[0,\infty)\times(k_{y,M},\infty)\}$. 
Based on the above analyses, the wave energy components within region $\Omega_1$ increase over time, whereas those within $\Omega_2$ remain finite.  {In $\Omega_2$, it always holds that $\Lambda(\bm k)>\bm k \cdot \bm U$, meaning that the resonance condition is not satisfied.  The wave dynamics in $\Omega_2$ can be analysed using a similar approach based on the weak wind condition discussed in \S~\ref{sec:4.3}.}  Hence, the function $E(k_y,t)$ can be expressed as:
\begin{align}
    E(k_y,t)&\leq 2\Vert \Phi_p (k_x^2+k_y^2)\Lambda^{-3} \Vert_{L^\infty_{x}}\left[M(k_y,t)+N(k_y,t)\right], &(k_x,k_y)\in \Omega_1,\label{eq:E2_gc1}\\
    E(k_y,t)&\leq 2\Vert \Phi_p (k_x^2+k_y^2)\Lambda^{-3} \Vert_{L^\infty_{x}}C_1, &(k_x,k_y)\in \Omega_2,
\end{align}
where $C_1$ is a constant that does not depend on time $t$.

Hence, the surface elevation variance $\langle\eta^2\rangle$ can be calculated from \eqref{eq:eta2_E} as follows:
\begin{align}
	\langle\eta^2\rangle
        \leq& 2\int_{-k_{y,M}}^{k_{y,M}} 
                \Vert \Phi_p(k_x^2+k_y^2)\Lambda^{-3}\Vert_{L_x^\infty}
                \left(M(k_y,t)+N(k_y,t)\right)
                \mathrm{d}k_y\notag\\
                &+2\left(\int_{-\infty}^{-k_{y,M}}+\int_{k_{y,M}}^{\infty} \right) \Vert \Phi_p(k_x^2+k_y^2)\Lambda^{-3}\Vert_{L_x^\infty} {C_1} \mathrm{d}k_y\notag\\
	\leq& 2\Vert\Phi_p(k_x^2+k_y^2)\Lambda^{-3}\mathbbm{1}_{|k_y|<k_{y,M}}M(k_y,t) 
                \Vert_{L_x^\infty L_y^1}\notag\\
	&+ 2\Vert\Phi_p(k_x^2+k_y^2)\Lambda^{-3}\mathbbm{1}_{|k_y|<k_{y,M}}N(k_y,t) 
                \Vert_{L_x^\infty L_y^1}+C_2,\label{eq:eta2_est_gravitycapillarywave}
\end{align}
where $\mathbbm{1}_{|k_y|<k_{y,M}}$ is the indicator function and $C_2$ is independent of time $t$.
The first term on the right-hand side of~\eqref{eq:eta2_est_gravitycapillarywave} represents the leading-order temporal increase in the wave energy, with the expression $M(k_y,t)$ explicitly given (see \cref{eq:expr_M}), and this term demonstrates the linear increases in the surface elevation variance over time. 
The residue term related to $N(k_y,t)$ represents the oscillatory behaviour of the wave energy evolution as well as the subdominant linear growth behaviour. The last term on the right-hand side of~\eqref{eq:eta2_est_gravitycapillarywave} represents to the contribution of the wave energy components that have wavenumbers for which the resonance does not occur.

In summary of~\S~\ref{sec4}, for gravity--capillary waves, we found that resonance occurs when the convection velocity of the air pressure fluctuations at the water surface is greater than the critical phase velocity of the water waves, which is equivalent to the condition for the nondimensional Froude number $\gamma_0>2$.  
When wind speeds fall below this threshold, the wind-generated surface elevation does not continually increase over time and instead saturates within a certain period.  When the resonance condition is satisfied (i.e., $\gamma_0>2$), our analyses demonstrate that the overall wave energy increases over time, and its leading-order component exhibits linear growth behaviour.

\section{Conclusions and discussions}\label{sec5}
In the present study, we revisit the important work of~\citet{phillips1957generation} on the resonance mechanism underlying wind-wave generation. 
In the initial stage of Phillips theory, the resonance between the phase velocity of the surface waves and the convection velocity of the air pressure fluctuations at the water surface induces the initial increase in the surface elevation variance. 
Wave energy components located on the resonance curve in the wavenumber space grow quadratically over time, while those far from the resonance curve show oscillatory behaviour. 
Phillips proposed that the total wave energy, i.e., the surface elevation variance, increases linearly over time. 
 {This initial stage of wind-wave generation process has been observed in both laboratory experiments and numerical simulations~\citep{zavadsky2017water,li2023direct}.}
While his theory has strong physical foundations, the original approach in \citet{phillips1957generation} lacks rigorous analyses, and a quantitative assessment of wave energy behaviour has been elusive.  
The present study introduces a new approach for evaluating the temporal growth of wave energy using a complex analysis method. 
Different from the Phillips approach, which approximates the integral of the wave energy components in resonance-curve based orthogonal coordinates, we directly perform the calculation in Cartesian coordinates in the wavenumber space.  
We employ the residue theorem to derive the results of the corresponding singular integral equations and explicitly obtain the leading-order term of the upper-bound solution of the wave energy, which increases linearly over time.

We highlight the critical role of surface tension in characterising the shape of resonance curves. 
To illustrate this concept, we first studied the evolution of gravity waves, for which surface tension is neglected. In this case, the resonance curve exhibits infinite length in the wavenumber space. 
The curve begins with a point on the $k_x$-axis and continues to infinity in the first quadrant of the wavenumber space. 
This resonance curve ensures that there exists a wavenumber satisfying the resonance condition for any given wavenumber $k_y$. 
We demonstrate that the integral of the wave energy components along the real $k_x$ axis for arbitrary values of $k_y$ can be obtained using a well-designed contour loop and the residue theorem, resulting in the summation of a linear function of time and an oscillatory function. 
This approach shows that the leading-order term of the upper-bound solution of the wave energy increases linearly over time for gravity waves in the initial stage of wind-wave generation, and the subdominant oscillatory term represents the fluctuating wave energy components.  

Different from gravity waves, the energy evolution of gravity--capillary waves exhibits distinct characteristics due to the inclusion of surface tension effects.
The inclusion of surface tension introduces a higher-order nonlinear term in the wave dispersion relation, leading to a resonance curve with distinct characteristics (see equation~\eqref{eq:cubic_eqn} and figure~\ref{phillips_fig}a). 
 {A critical value exists for the convection velocity of the air pressure fluctuations at the water surface, which equals the minimum phase velocity of gravity--capillary water waves}. 
When the convection velocity is below this threshold, no resonance occurs for any wavenumber, causing the wave energy to saturate. 
We propose a quantitative estimation of the wave energy~\eqref{eq:weak_est_G2} and evaluate its dependency on the convection velocity of the air pressure fluctuations. 
This approach successfully extends the original analysis of the wave energy under weak wind conditions by~\citet{phillips1957generation}, which considered only the case of zero convection velocity for the air pressure fluctuations.  
When the convection velocity exceeds the critical value, resonance occurs when the wavenumbers are on the resonance curve. 
Under such strong wind conditions, the resonance curve exhibits a finite length in the wavenumber space. 
This curve starts at one point on the positive $k_x$-axis, extends to the first quadrant of the wavenumber space, and ends at another point on the positive $k_x$-axis, as depicted in figure~\ref{phillips_fig}($a$). 
These characteristics of the resonance curve under strong wind conditions ensure that the integration of wave energy components with respect to the $k_x$ variable increases over time when the magnitude of $k_y$ is less than a critical value to satisfy the resonance condition. 
Using the complex analysis method, we explicitly obtained the leading-order term of the upper-bound solution of the wave energy, which increases linearly over time. {This term represents the primary contribution compared to the subdominant term in most cases.}
The subdominant term, which was obtained by evaluating a specific integral path, shows a similar linear growth behaviour as well as oscillatory phenomenon over time. 
When the spanwise wavenumber $k_y$ exceeds the critical value, no resonance occurs when integrating the wave energy components with respect to the $k_x$ variable. 
Consequently, the wave energy in this wavenumber region is saturated instead of increasing over time. 

We remark that the present study focuses on analytical solutions of the upper bound of wave energy in the initial stage of wind-wave generation.  
Because of the turbulent nature of airflows, directly incorporating air pressure fluctuations into the complex analysis method is infeasible at present due to the lack of an analytical model for turbulent pressure, as this method requires the variables to be analytic functions.  
Therefore, we use inequality estimations to obtain an upper-bound estimation of wave energy growth.  
Through the complex analysis, we obtain analytical results that demonstrate the leading-order term and subdominant term of the upper-bound solution of the wave energy growth for gravity and gravity--capillary waves for the first time.  
The quantitative solutions provide an insightful analytical perspective on wind-wave generation process.  
The theoretical framework developed in the present study, which extends the seminal resonance theory proposed by Phillips~(\citeyear{phillips1957generation}), will be useful for the further model development in the future, for which modelling research on turbulent pressure fluctuations in the turbulent boundary layer research community will be helpful.  
 {We also note that the present study does not consider water dissipation, which is consistent with Phillips theory.  The influence of water dissipation on wave growth can be further incorporated based on the method introduced in~\citet{perrard2019turbulent}, which is of interest for future research that further considers the effects of wind-induced sheared water currents~\citep[e.g.,][]{nove2020effect,geva2022excitation}.}

\vspace{20pt}

\noindent{\bf{Acknowledgements.}}
This work was supported by the Office of Naval Research.  A portion of T.L.’s work was performed under the auspices of the U.S. Department of Energy by Lawrence Livermore National Laboratory under Contract DE-AC52-07NA27344 with IM release number LLNL-JRNL-859400 and was supported by the LLNL-LDRD Program under Project No. 25-LW-087. We thank the three reviewers for their insightful and constructive comments.
	
\bibliographystyle{jfm}

\end{document}